\documentclass[preprintnumbers,floatfix,letterpaper,aps,prd,nofootinbib,onecolumn]{revtex4-2}
\usepackage{amsmath,amsfonts,latexsym,amssymb,graphicx,graphics,epsfig,subfigure,color,makeidx,array}
\usepackage{xcolor,diagbox}
\usepackage{multirow}
\usepackage[colorlinks,linkcolor=blue,anchorcolor=blue,citecolor=blue,urlcolor=blue]{hyperref}
\usepackage{mathrsfs}
\usepackage{amsmath}
\usepackage{bm,graphicx,dcolumn,epstopdf,epsf,latexsym,mathbbol, amssymb,amsmath,color,slashed, mathrsfs,mathcomp,simplewick}
\usepackage{makecell}

\newcommand{\bq}{\begin{equation}}
\newcommand{\eq}{\end{equation}}
\newcommand{\bqn}{\begin{eqnarray}}
\newcommand{\eqn}{\end{eqnarray}}
\newcommand{\nb}{\nonumber}
\newcommand{\lb}{\label}
\newcommand{\f}{\frac} 

\newcommand{\tx}{\text}

\newcommand{\lf}{\left}
\newcommand{\rt}{\right}

\begin{document}

\title{Constraints on quantum Oppenheimer-Snyder black holes with eccentric extreme mass-ratio inspirals}

\author{Sen Yang${}^{a, b}$}
\email{120220908881@lzu.edu.cn}

\author{Yu-Peng Zhang${}^{a, b}$}
\email{zhangyupeng@lzu.edu.cn}

\author{Li Zhao${}^{a, b}$}
\email{lizhao@lzu.edu.cn, corresponding author}

\author{Yu-Xiao Liu${}^{a, b}$}
\email{liuyx@lzu.edu.cn, corresponding author}

\affiliation{
${}^{a}$ Lanzhou Center for Theoretical Physics, \\
Key Laboratory of Theoretical Physics of Gansu Province,\\
Key Laboratory for Quantum Theory and Applications of MoE,\\
Gansu Provincial Research Center for Basic Disciplines of Quantum Physics\\
Lanzhou University, Lanzhou 730000, China\\
${}^{b}$ Institute of Theoretical Physics $\&$ Research Center of Gravitation,\\
School of Physical Science and Technology,\\Lanzhou University, Lanzhou 730000, China\\}
\date{\today}

\begin{abstract}

We investigate the potential of extreme mass-ratio inspirals to constrain quantum Oppenheimer–Snyder black holes within the framework of loop quantum gravity. We consider a stellar-mass object orbiting a supermassive Oppenheimer-Snyder black hole in an equatorial eccentric trajectory. To explore the dynamical behavior of the system, we analyze its orbital evolution under gravitational radiation within the adiabatic approximation and the mass-quadrupole formula for different initial orbital configurations. Our results show that the quantum correction parameter $\hat{\alpha}$ slows down the evolution of the orbital semi-latus rectum and eccentricity. We then employ the numerical kludge method to generate the corresponding time-domain gravitational waveforms. To assess detectability, we include Doppler modulation due to the motion of space-based detectors and compute the frequency-domain characteristic strain. By evaluating mismatches between response signals for different values of $\hat{\alpha}$, we show that even small corrections $( \hat{\alpha} \sim 10^{-5})$ produce distinguishable effects. Our analysis suggests that future space-based detectors such as LISA can probe quantum gravitational corrections in the strong-field regime and place constraints significantly stronger than those from black hole shadow observations.

\end{abstract}


\maketitle

\section{Introduction}
\label{Introduction}
\renewcommand{\theequation}{1.\arabic{equation}} 
\setcounter{equation}{0}

Extreme mass-ratio inspirals (EMRIs), where a stellar-mass object orbits and gradually fall into a supermassive black hole, are promising sources for future space-based gravitational wave detectors, such as the Laser Interferometer Space Antenna (LISA) \cite{LISA:2017pwj}, Taiji \cite{Hu:2017mde}, TianQin~\cite{TianQin:2015yph}, and the DECi-hertz Gravitational-wave Observatory (DECIGO)~\cite{Musha:2017usi}. In an EMRI system, the smaller object loses energy and orbital angular momentum as it spirals inward, due to the emission of gravitational waves ~\cite{Apostolatos:1993nu, Tanaka:1993pu, Cutler:1994pb}. Unlike the short-duration signals from the merger of stellar-mass binaries, the gravitational waves from EMRIs can last for years within the sensitive frequency band of these detectors~\cite{Amaro-Seoane:2012lgq}. These long-lasting signals carry detailed information about the supermassive black hole, the structure of the surrounding spacetime, and the nature of gravity itself~\cite{Hughes:2000ssa}. Therefore, EMRIs provide a unique opportunity to study black hole physics and to test gravitational theories in the strong-field regime~\cite{Glampedakis:2005hs, Gair:2012nm, Barausse:2020rsu, LISA:2022yao, GWSTQLISA, Babak:2017tow}.

The study of EMRIs has advanced significantly with progress in both theoretical models and computational methods. Black hole perturbation theory is a widely used method, in which gravitational waves are treated as perturbations of the supermassive black hole spacetime, and the smaller object acts as the source of these perturbations ~\cite{Zerilli:1970wzz, Davis:1971gg, Davis:1972ud}. This approach led to the development of the gravitational self-force framework~\cite{Quinn:1996am, Lousto:1999za, Barack:2002mh}. As computational resources improved, new methods such as the effective one-body approach were introduced~\cite{Buonanno:1998gg, Taracchini:2013rva, Yunes:2009ef, Zhang:2021fgy}. The effective one-body method combines analytical and numerical techniques to capture higher-order relativistic effects. However, simulating the full dynamics of EMRIs remains computationally demanding. This challenge motivated the development of analytical and numerical kludge (NK) models~\cite{Barack:2003fp, Babak:2006uv, Sopuerta:2011te, Chua:2017ujo, Liu:2020ghq}. The NK method combines the exact relativistic orbit of the smaller object with an approximate formula for the emitted gravitational waves~\cite{Babak:2006uv}. It treats the radiation as linear perturbations on flat spacetime, enabling faster computation than full numerical relativity. Although less accurate than full simulations, the NK method offers a good balance between efficiency and precision, making it a widely used tool in EMRI gravitational wave research.

The advancement of modeling techniques has opened new avenues for probing black hole physics through gravitational waves from EMRIs. Recent developments in loop quantum gravity suggest that quantum corrections may subtly modify the geometry of black hole spacetimes~\cite{Ashtekar:2004eh, Perez:2017cmj, Zhang:2023yps}. Although these corrections are typically small, they can influence the orbital dynamics of EMRIs and affect the phase evolution of the emitted gravitational waves. Several studies have investigated how such quantum effects could appear in the gravitational waveforms of EMRIs ~\cite{Tu:2023xab, Liu:2024qci, Yang:2024lmj, Jiang:2024cpe, Zhang:2024csc, Zi:2024jla, Fu:2024cfk, Yang:2024cnd}. These works indicate that quantum modifications may lead to detectable phase shifts in the waveforms, offering a promising avenue to place constraints on quantum gravity parameters using future space-based gravitational wave observations.

Gravitational waveforms from periodic orbits around the quantum Oppenheimer-Snyder black hole have previously been investigated in Ref.~\cite{Yang:2024lmj}. However, that study was limited to short-duration orbital motion and did not account for the backreaction effects of gravitational radiation. In this paper, we extend the analysis to long-duration EMRIs in eccentric orbits around the quantum Oppenheimer–Snyder black hole, accounting for energy and angular momentum losses due to gravitational wave emission. We model the secular orbital evolution of a stellar-mass compact object orbiting a supermassive black hole by applying the adiabatic approximation together with the mass-quadrupole radiation formula. The energy and angular momentum losses due to gravitational wave emission are computed numerically. It allows us to track the evolution of the orbital elements (the semi-latus rectum and eccentricity) over time. Using the computed orbits, we generate time-domain gravitational waveforms with the numerical kludge method. To evaluate the observational implications, we incorporate Doppler modulation caused by the motion of a LISA-like space-based detector and simulate its response to the gravitational signals. We then analyze the frequency-domain characteristics of the waveforms. By fixing the signal-to-noise ratio (SNR) and computing waveform mismatches for different values of $\hat{\alpha}$, we assess the detectability of quantum corrections.

This paper is organized as follows. In Sec.~\ref{sec2}, we study the orbital evolution of a small object around a quantum Oppenheimer-Snyder black hole under the influence of gravitational radiation. In Sec.~\ref{sec3}, we compute the corresponding gravitational waveforms. In Sec.~\ref{sec4}, we analyze the detector response to these signals and explore the potential of future space-based detectors to constrain the quantum correction parameter. Finally, the conclusions and discussions of this work are given in Sec.~\ref{sec5}. Throughout this paper, we adopt the geometrized unit system with $G = c = 1$.

\section{background and Orbital evolution} 
\label{sec2}
\renewcommand{\theequation}{2.\arabic{equation}} 
\setcounter{equation}{0}

The classical Oppenheimer-Snyder model describes the collapse of a uniformly distributed matter-filled universe into a black hole, ultimately forming a gravitational singularity~\cite{Oppenheimer:1939ue}. By incorporating quantum gravity effects, a quantum-corrected version of the Oppenheimer–Snyder black hole was proposed in Ref.~\cite{Lewandowski:2022zce}. The metric of this quantum Oppenheimer-Snyder black hole is given by
\bqn\lb{metric}
ds^2 = - f(r) dt^2 + f(r)^{-1} dr^2 + r^2 d \theta^2 + r^2 \sin^2 \theta d \phi^2 ,
\eqn 
where
\bqn
f(r) =  1 - \f{2 M}{r} + \f{\alpha M^2}{r^4}.
\eqn 
Here, $M$ is the Arnowitt–Deser–Misner mass of the black hole, and $\alpha$ is the parameter characterizing quantum corrections. Equation~\eqref{metric} reduces to the metric of a Schwarzschild black hole when the parameter $\alpha$ vanishes. For convenience, we define a dimensionless quantum correction parameter $ \hat{\alpha} = \alpha/M^2$, which is used throughout this work. Previous studies~\cite{Zhao:2024elr, Ali:2024ssf, Vachher:2024ait} have constrained the parameter $\hat{\alpha}$ using observations of the black holes' shadows of M87$^\ast$ and Sgr A$^\ast$, as well as strong gravitational lensing effects, and found that the upper bound on $\hat{\alpha}$ is $1.4087$.

A central aspect of the EMRI problem is studying the orbital evolution of the smaller object under the influence of gravitational radiation. The analytical kludge approach is a widely used and effective method for modeling the orbital evolution of EMRIs under the influence of gravitational radiation~\cite{Barack:2003fp, Sopuerta:2011te, Chua:2017ujo, Liu:2020ghq}. However, this approach approximates the small object's trajectory by using Keplerian orbits. In this work, we consider a
stellar-mass object orbiting a supermassive Oppenheimer-Snyder black hole on the equatorial plane. The equations of motion for the object on the equatorial plane are shown in Appendix~\ref{AppA}. We adopt a numerical scheme based on the adiabatic approximation to study the orbital evolution of this EMRI system under the influence of gravitational radiation.

For a given initial pericenter distance $r_{p}$ and eccentricity $e_0$ of the smaller object's orbit, we first calculate the specific energy $E_0$ and specific orbital angular momentum $L_0$ of the orbit by using Eq.~\eqref{dotr}, with the conditions that the radial velocity of the object vanishes at both the pericenter and apocenter. In the initial stage, we ignore the effect of gravitational radiation and evolve the orbit for ten orbital periods purely using geodesic equations of motion~\eqref{dott},~\eqref{dotphi}, and~\eqref{dotr}. During this phase, we use the quadrupole approximation to compute and record the energy and angular momentum fluxes of gravitational radiation at each time step as
\bqn\lb{fluex-1}
\f{dE_\text{GW}}{dt} = \f{1}{5}  \dddot{I}_{ij} \dddot{I}_{ij},~~~~\f{dL_\text{GW}^i}{dt} = \f{2}{5} \epsilon ^{ikl} \ddot{I}_{ka} \dddot{I}_{la},
\eqn
where $\epsilon ^{ikl}$ is the three-dimensional Levi-Civita symbol, and the symmetric and trace-free (STF) mass quadrupole of the smaller object can be defined as \cite{Thorne:1980ru}
\bqn\lb{Iij}
I^{ij} = \left[ \int d^3x T^{tt}(t, x^l) x^i x^j  \right]^{(\text{STF})},
\eqn 
where $T^{tt}$
is the $tt$-component of the stress-energy tensor for the object. These recorded fluxes are used to construct a time-averaged radiation window over one initial orbital period as~\cite{Maggiore:2007ulw}
\bqn
\left\langle \frac{dX}{dt} \right\rangle = \frac{1}{T_{\text{ave}}} \int \left[ \frac{dX}{dt}(t) \right] dt = \frac{1}{N_{\text{ave}}} \sum_{i=1}^{N_{\text{ave}}} \frac{dX}{dt}(t),
\eqn
where $T_{\text{ave}}$ is the time of the initial orbital period, and $N_{\text{ave}}$ is the corresponding step number.  The fluxes of orbital energy and angular momentum are
\bqn
\frac{d E}{d t} = - \left\langle \frac{d E_\text{GW}}{dt} \right\rangle,~~~~\frac{d L^z}{d t} = - \left\langle \frac{d L_\text{GW}}{dt} \right\rangle.
\eqn
After the initial period evolution, we begin updating the orbital energy and angular momentum at each time step with the averaged fluxes in the radiation window 
\bqn
E_{i+1} = E_{i} + \left. \frac{d E}{d t} \right|_{i} \Delta t,~~~~L^z_{i+1} = L^z_{i} + \left. \frac{d L^z}{d t} \right|_{i} \Delta t,
\eqn
where $\Delta t$ is the time step for numerical evolution. As the orbit evolves, we dynamically update the radiation window by incorporating newly computed fluxes and discarding the oldest ones. The averaged fluxes are then used to calculate the change in the orbit's energy and angular momentum due to gravitational wave emission. These values are updated step-by-step to evolve the orbit under radiation reaction. The orbital evolution is terminated once the radial coordinate of the orbit becomes less than or equal to the radius of the innermost stable circular orbit of the quantum black hole.

\begin{figure}[!h]
	\centering
    {\includegraphics[scale =0.24]{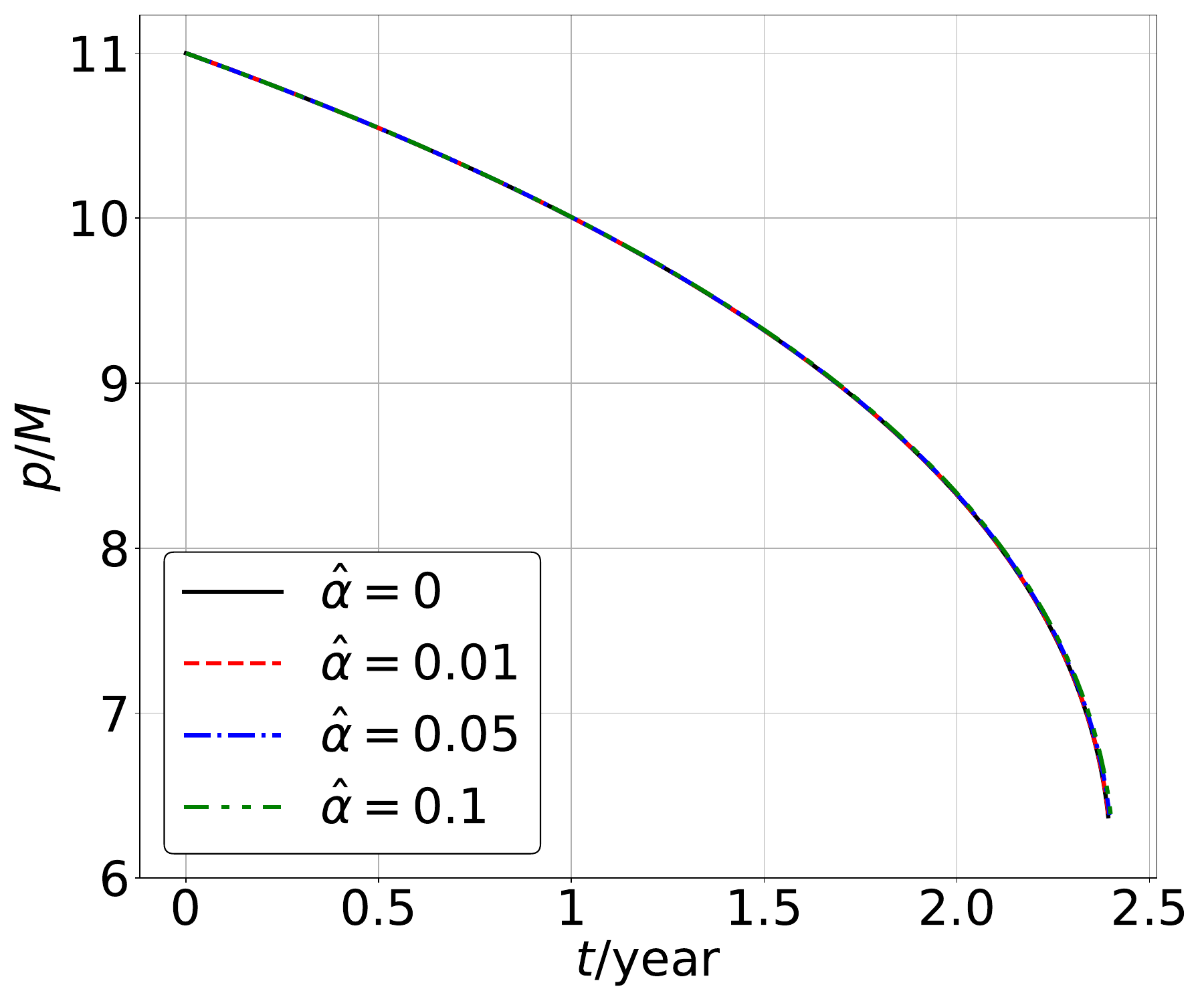}\lb{}}
   {\includegraphics[scale =0.24]{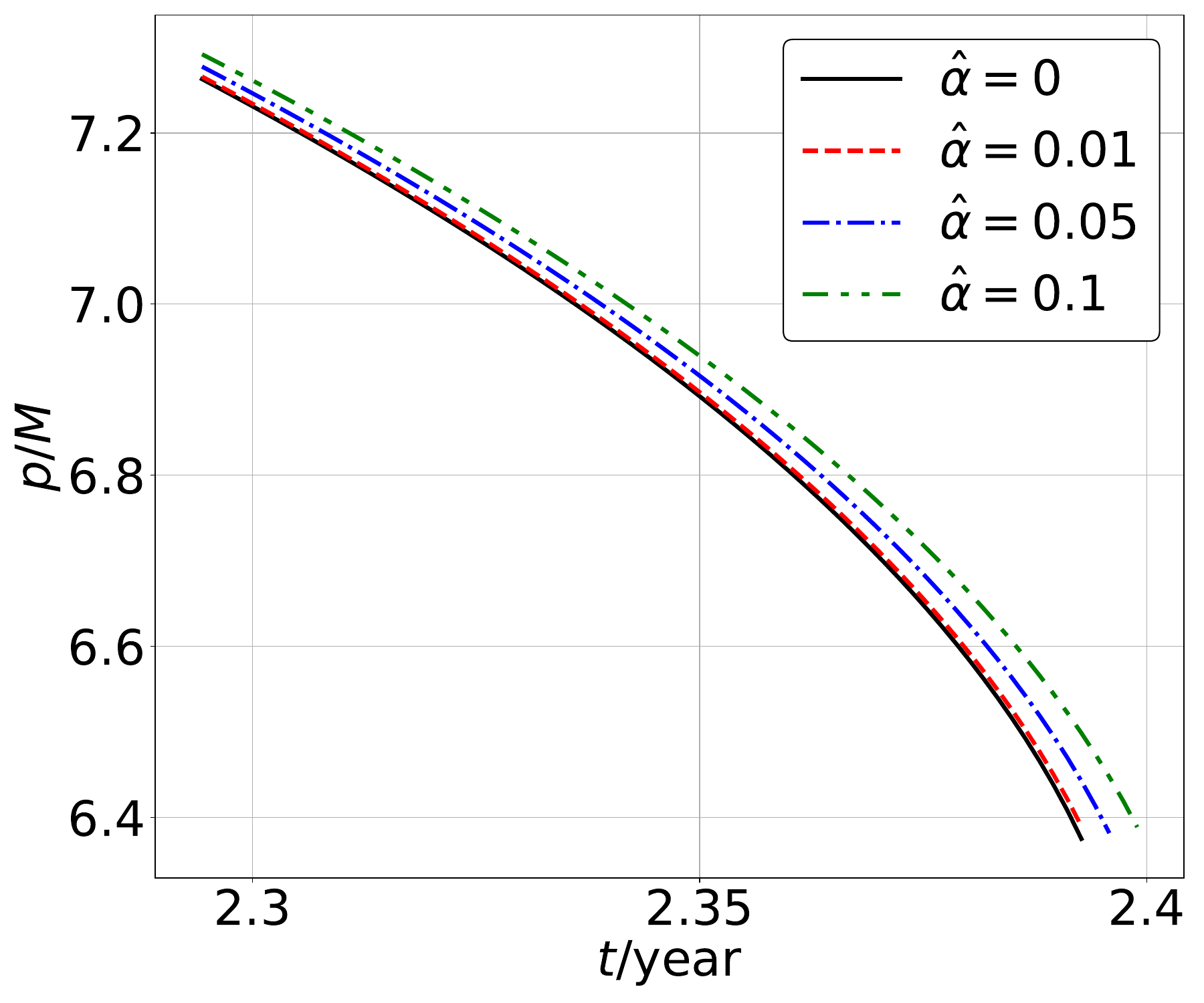}\lb{}}
  {\includegraphics[scale =0.24]{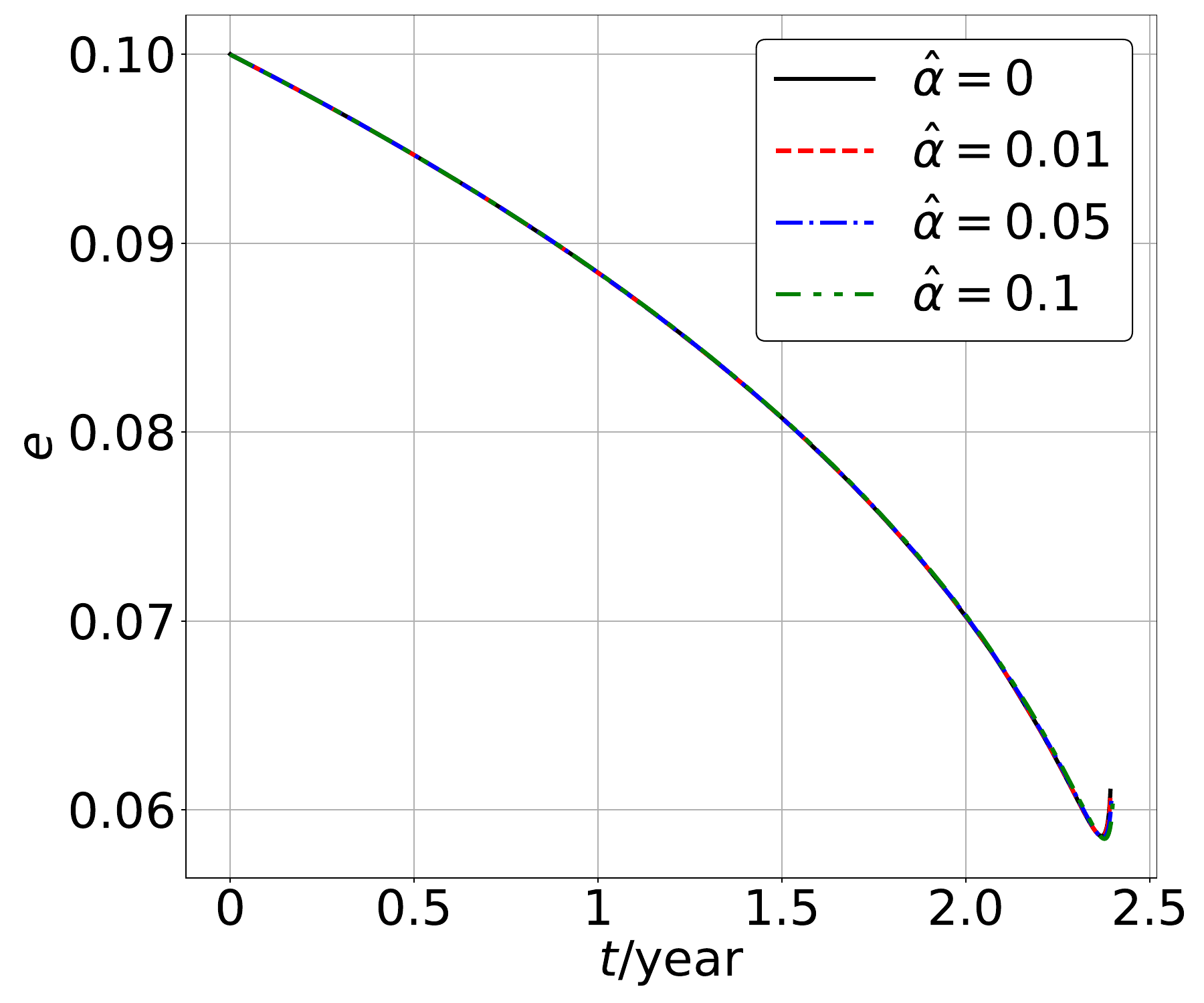}\lb{}}
    {\includegraphics[scale =0.24]{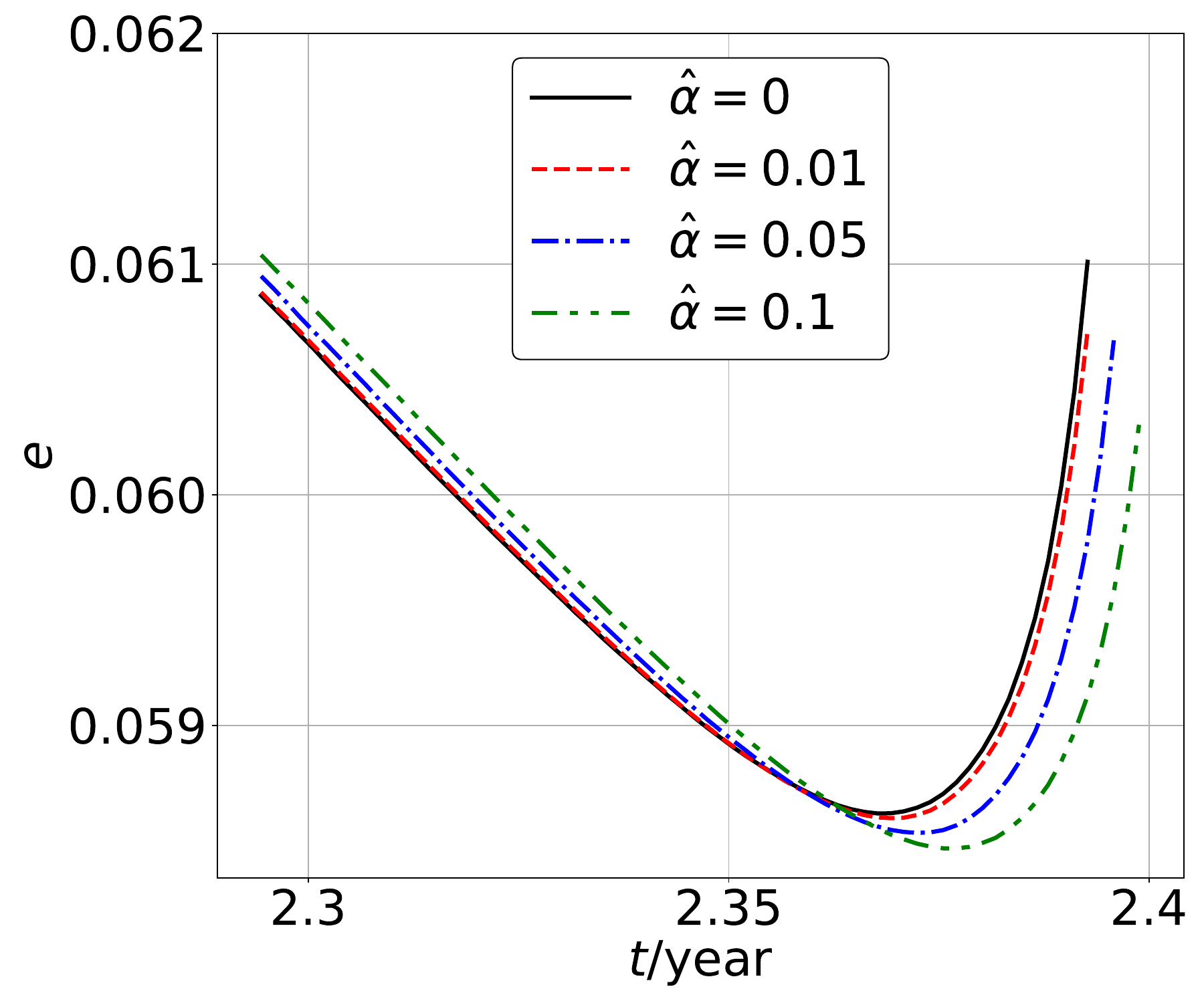}\lb{}}
	\caption{The evolutions of the semi-latus rectum $p$ and eccentricity $e$ of the smaller object’s orbit with initial $r_p = 10 M$ and $e=0.1$ around the supermassive quantum Oppenheimer-Snyder black hole.}
	\label{Evolutions-of-p-e-1}
\end{figure}

\begin{figure}[!h]
	\centering
    {\includegraphics[scale =0.24]{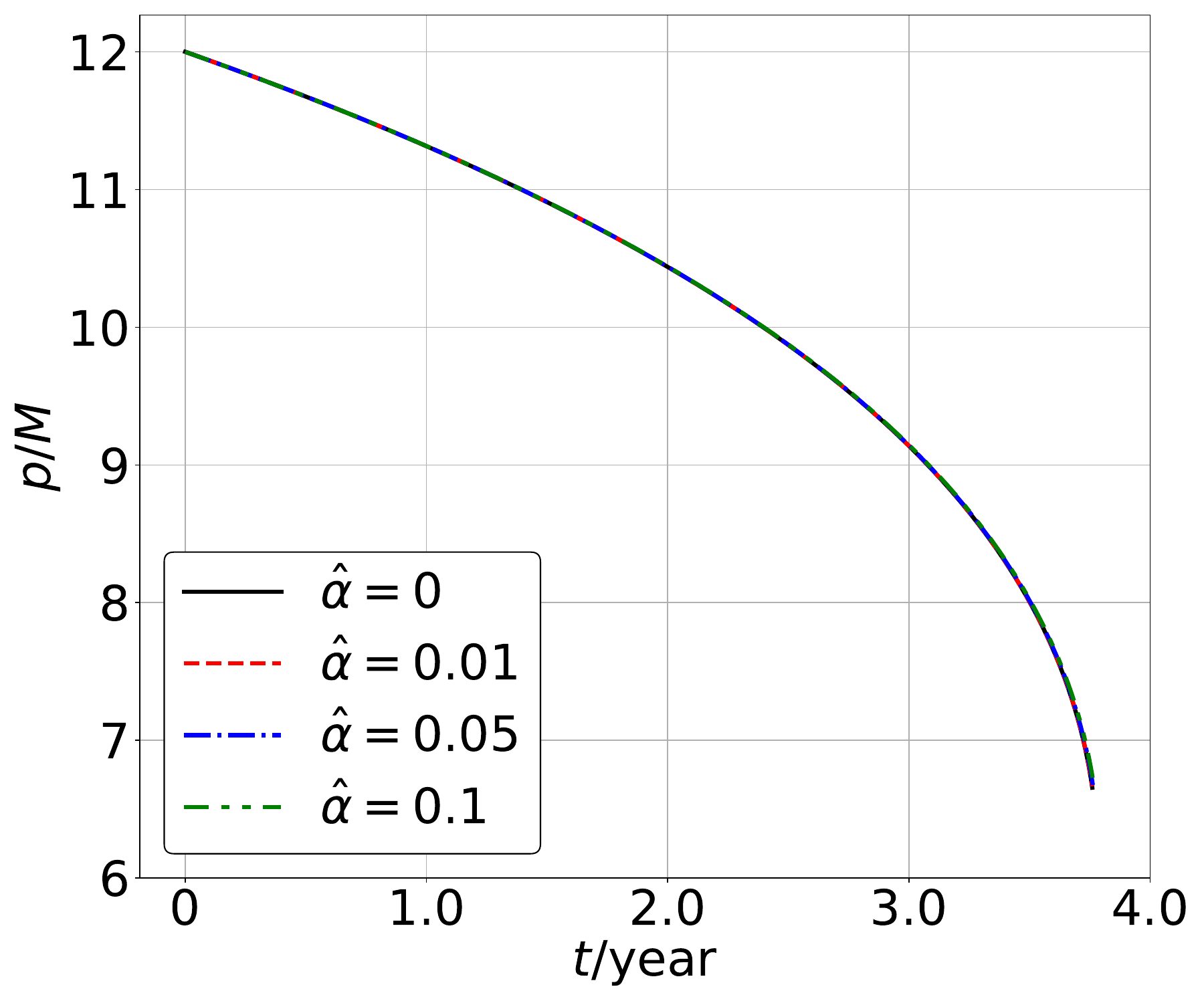}\lb{}}
    {\includegraphics[scale =0.24]{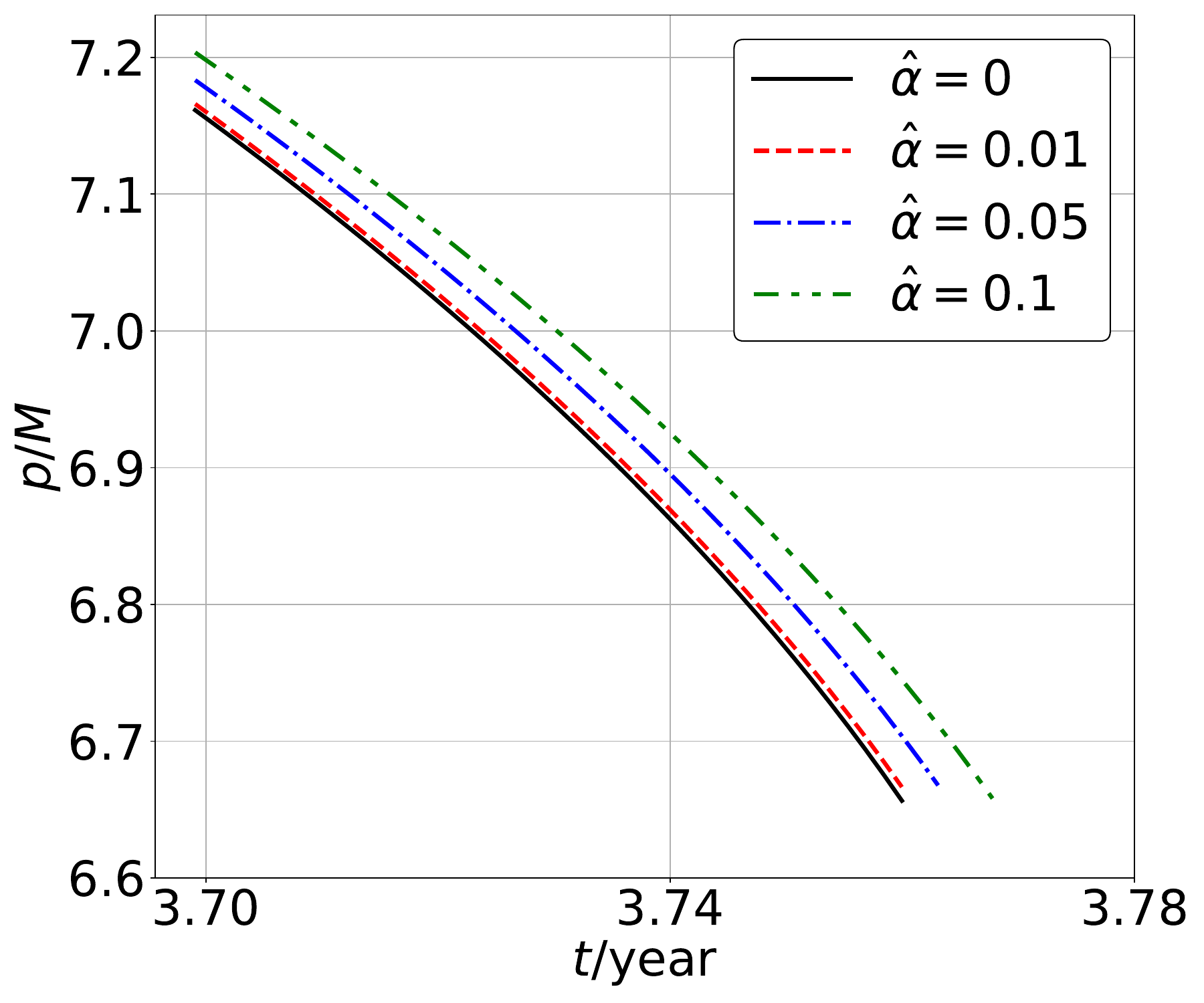}\lb{}}
    {\includegraphics[scale =0.24]{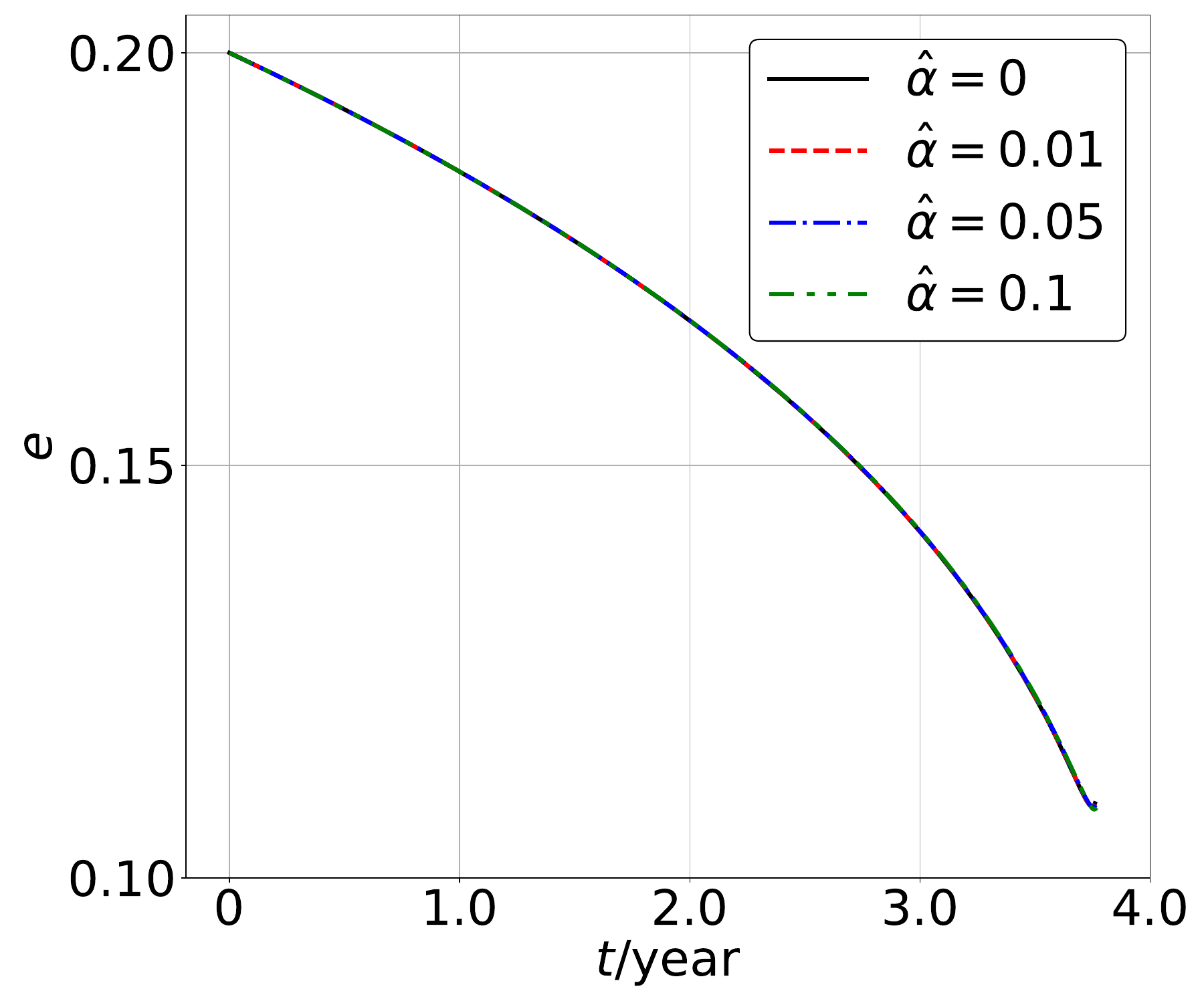}\lb{}}
    {\includegraphics[scale =0.24]{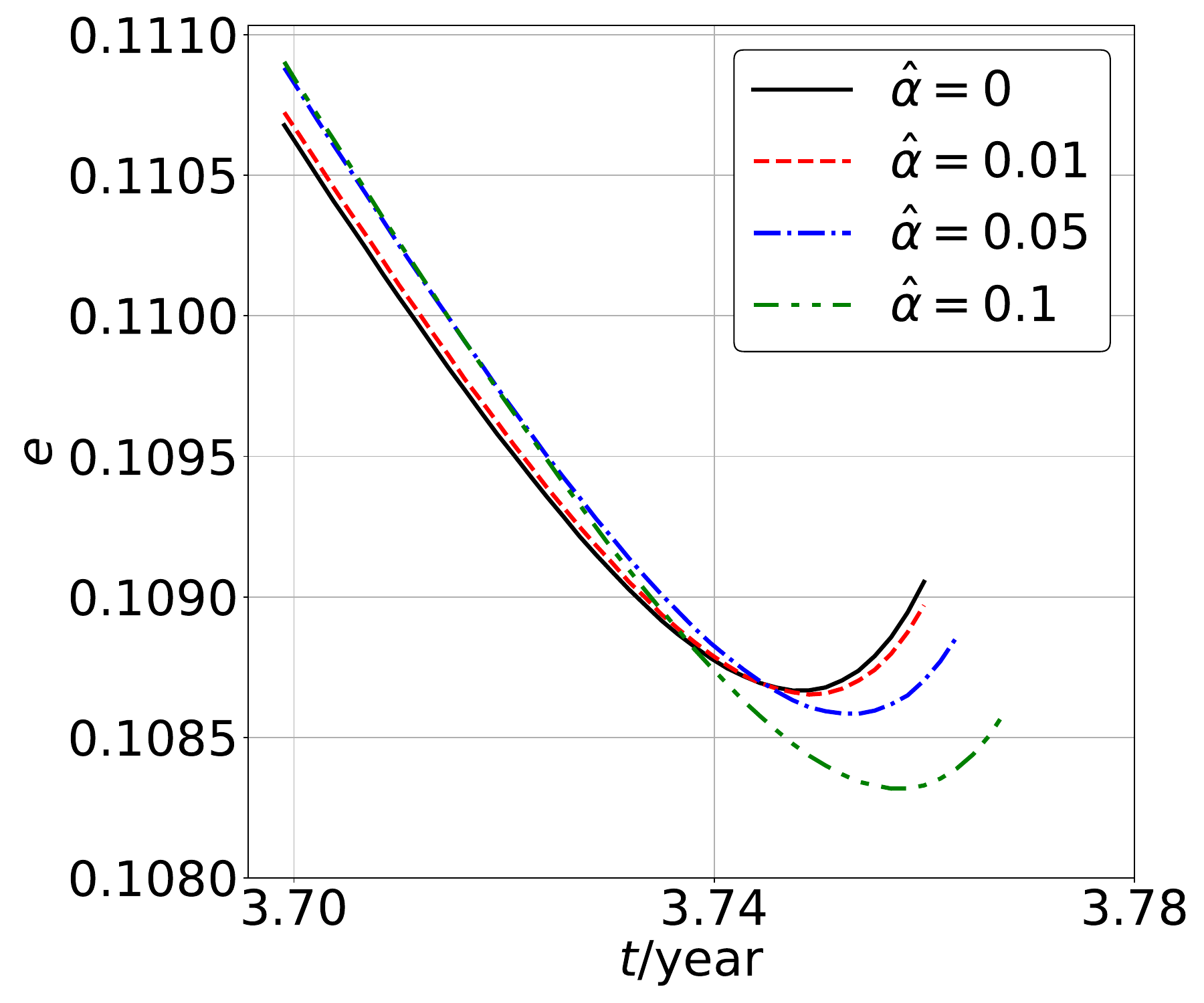}\lb{}}
	\caption{The evolutions of the semi-latus rectum $p$ and eccentricity $e$ of the smaller object’s orbit with initial $r_p = 10 M$ and $e=0.2$ around the supermassive quantum Oppenheimer-Snyder black hole.}
	\label{Evolutions-of-p-e-2}
\end{figure}

Based on the numerical algorithm described above, we perform the orbital evolution of a small object under the influence of gravitational radiation. We set the mass of the supermassive quantum Oppenheimer-Snyder black hole to be $M=10^6 M_\odot$, and the mass of the small object to be $ m=10 M_\odot$. The initial orbit is specified by a pericenter distance $r_p=10M$ and orbital eccentricities 
$e_0 = \{0.1,~0.2\}$. The initial position of the small object is set at $r_0 = r_p$ and $\phi_0 = 0$. Under these initial conditions, we numerically evolve the orbit with energy and orbital angular momentum loss due to gravitational radiation. To better characterize the orbital evolution of the small object, we calculate the semi-latus rectum and eccentricity at each moment during the evolution by using Eqs.~\eqref{p} and~\eqref{e}, respectively. We plot the results in Figs.~\ref{Evolutions-of-p-e-1} and \ref{Evolutions-of-p-e-2}.

Figures~\ref{Evolutions-of-p-e-1} and~\ref{Evolutions-of-p-e-2} show the evolution of the semi-latus rectum
$p$ and eccentricity $e$ for two types of orbits. In each case, the results are presented for the Schwarzschild black hole and for the quantum-corrected black hole with various values of the parameter $\hat{\alpha}$. One can find that, for all initial conditions, $p$ decreases monotonically and at an accelerating rate over time due to gravitational radiation, which is consistent with the expected inspiral behavior. When the initial eccentricity is $0.1$ or $0.2$, the eccentricity decreases gradually at first, and then increases again during the late stages of evolution. This behavior results from the effects of gravitational radiation~\cite{Cutler:1994pb}. The quantum correction slightly alters the rate of orbital evolution. In particular, larger values of $\hat{\alpha}$ lead to a slower decrease in $p$ and $e$ at late times, as shown more clearly in the zoomed-in plots on the right panels. The deviation from the general relativity trajectory becomes more pronounced as $\hat{\alpha}$ increases. 

\begin{figure}[!t]
	\centering
    \subfigure[$e_0 = 0.1$]{\includegraphics[scale =0.24]{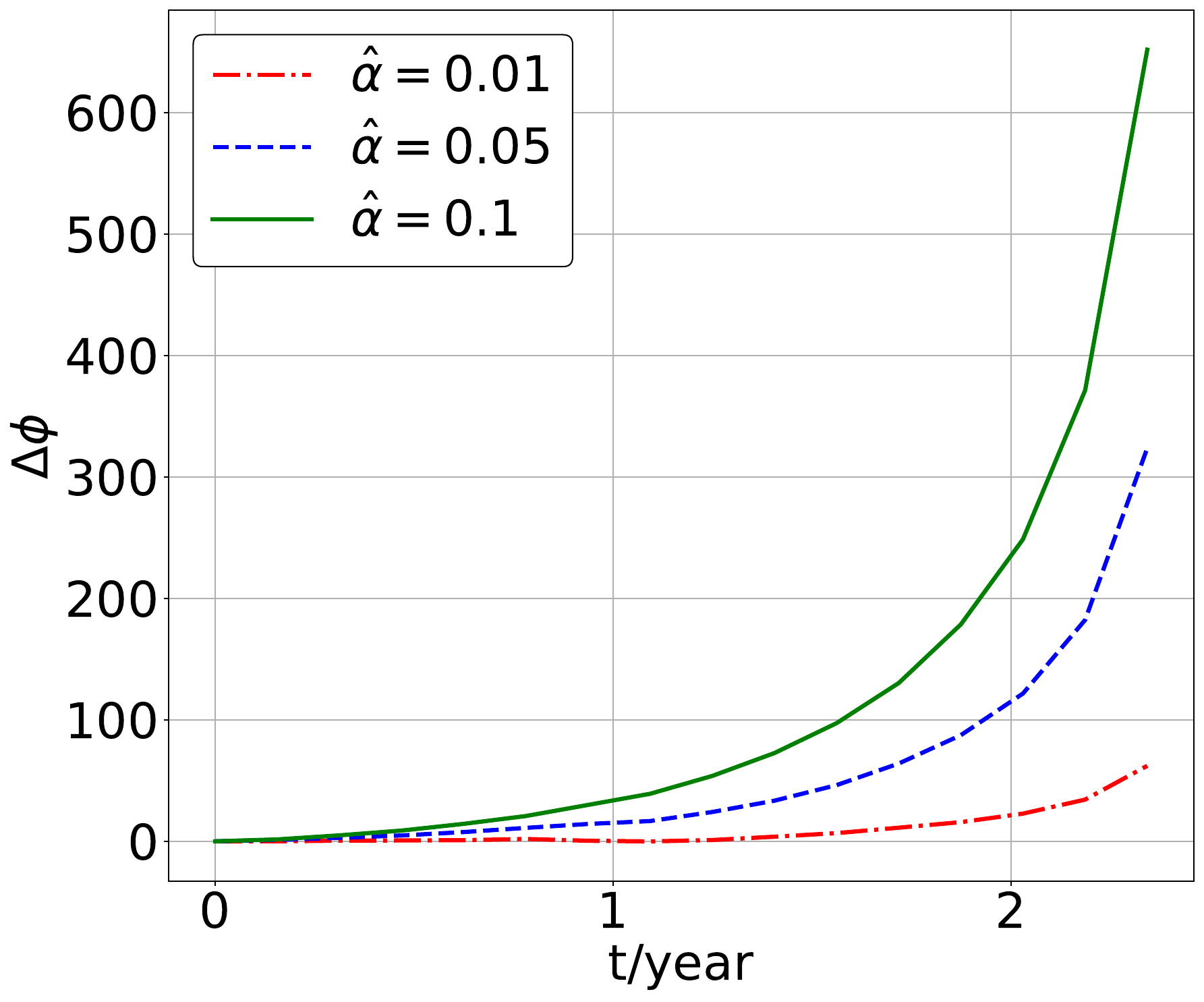}}
    \subfigure[$e_0 = 0.2$]{\includegraphics[scale =0.24]{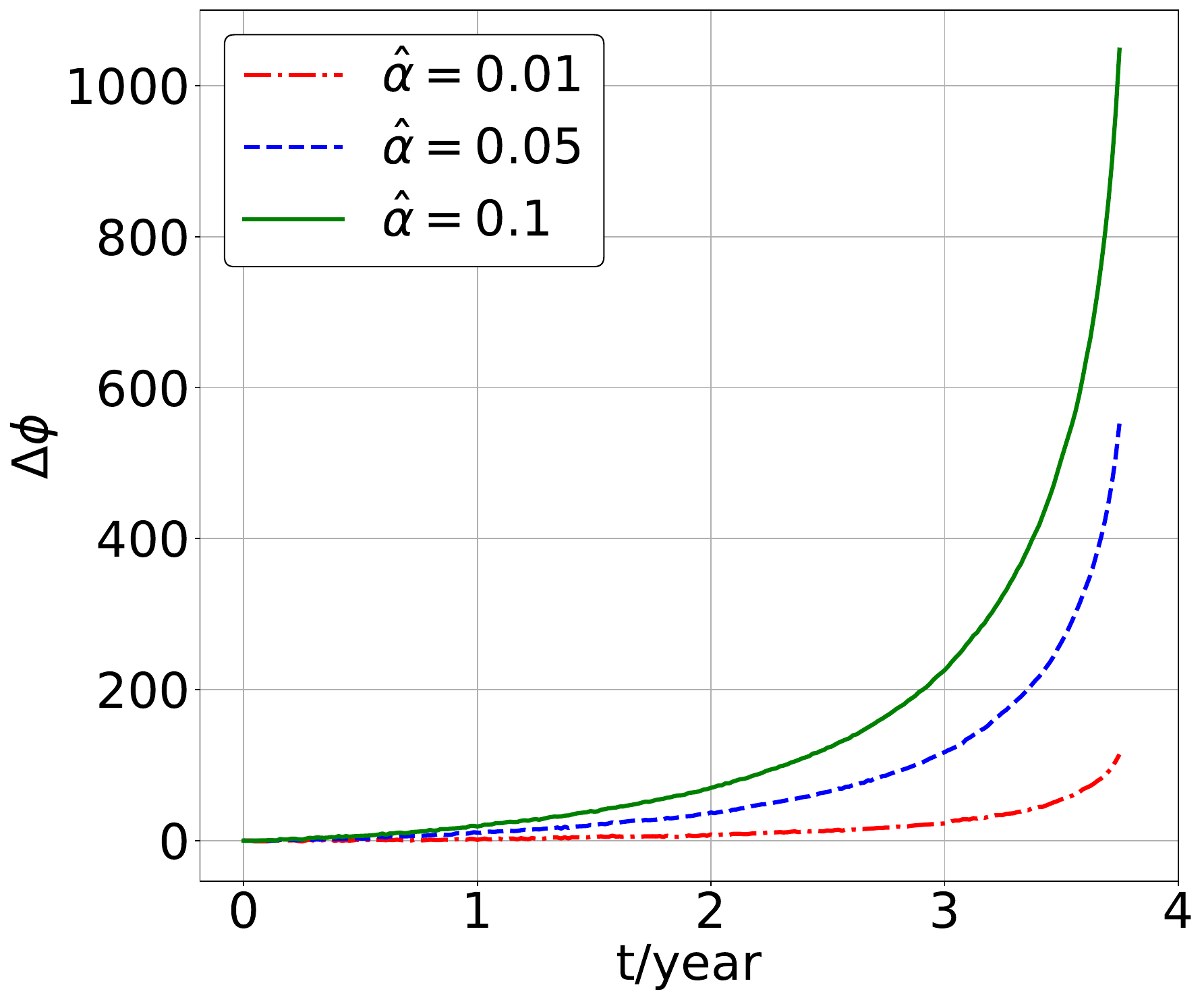}}
	\caption{Dephasing of the orbital phase for different values of the quantum correction parameter $\hat{\alpha}$.}
	\label{phase}
\end{figure}

To gain a clearer understanding of how the parameter $\hat{\alpha}$ affects the orbital phase evolution, we introduce the dephasing as
\bqn\lb{phase-Eq}
\Delta \phi = \phi(\hat{\alpha}) -\phi_0, 
\eqn 
where $\phi_0$ denotes the orbital phase in the case of a supermassive Schwarzschild black hole. Using Eq.~\eqref{phase-Eq}, we compute the dephasing for several values of 
$\hat{\alpha}$. The results are shown in Fig.~\ref{phase}. The analysis demonstrates that the parameter $\hat{\alpha}$ generates an advance in the orbital phase, and the magnitude of this phase deviation increases with the parameter $\hat{\alpha}$. Furthermore, the phase advancement accumulates over the inspiral. This is consistent with the conclusions reported in Ref.~\cite{Yang:2024lmj}. These results suggest that although the influence of quantum gravity effects may be negligible on short timescales, their persistent contribution may accumulate to produce detectable modifications in the orbital dynamics of EMRIs. Since the orbital phase evolution directly determines the gravitational waveform, such modifications in the orbital dynamics will inevitably leave imprints on the waveforms of EMRIs. It should also be emphasized that the parameter values adopted here are deliberately chosen to be much larger than realistic estimates, in order to illustrate the qualitative role of quantum corrections more clearly.

It is worth noting that the quadrupole formula and the numerical kludge method used in this work offer an efficient way to capture the leading-order gravitational radiation and the overall waveform morphology for a proof-of-concept study. However, in the strong-field region near the central supermassive black hole, higher-order relativistic effects -- in particular the gravitational self-force~\cite{Quinn:1996am, Lousto:1999za, Barack:2002mh} and post-adiabatic corrections~\cite{Hinderer:2008dm, Flanagan:2012kg, LISA:2022kgy} -- can modify the orbital evolution and, consequently, the accumulated waveform phase. Although these effects become most significant only in the strong-field stage, this stage occupies a relatively small fraction of the entire EMRI evolution, which typically spans a very long duration. Even so, these corrections encode the strong-field effects of the central supermassive black hole, can lead to cumulative dephasing, and may influence quantitative parameter estimation. For this reason, we interpret our numerical results as indicating qualitative trends and order-of-magnitude behavior. A full quantitative assessment using gravitational self-force formalism represents a promising avenue for further study. 

\section{gravitational waveforms}\lb{sec3}
\renewcommand{\theequation}{3.\arabic{equation}}
\setcounter{equation}{0}

In this section, we compute the gravitational waveforms emitted by the small object during its inspiral around the quantum-corrected black hole. The waveform is determined by the motion of the small object, which we have obtained in Section 2. To calculate the gravitational waveform, we use the quadrupole approximation~\cite{Thorne:1980ru}. The quadrupole formula provides a simple and efficient way to model the waveform from a slowly evolving EMRI system. We adopt the numerical kludge approach~\cite{Babak:2006uv}, which combines accurate orbital dynamics with a simple waveform generation scheme based on flat-spacetime radiation formulas. In this method, the small object’s trajectory is treated as a geodesic that evolves slowly due to energy and angular momentum losses. At each step of the evolution, we compute the waveform using the quadrupole formula and the instantaneous position and velocity of the object.

Under the mass–quadrupole approximation, the metric perturbations in the transverse-traceless gauge for EMRIs are \cite{Thorne:1980ru, gravity-book}
\bqn\lb{metric perturbations}
h_{ij} = \f{2}{D_\tx{L}} \frac{d^2 I_{ij}}{dt^2},
\eqn 
where $D_\tx{L}$ is the luminosity distance. The gravitational-wave polarizations are obtained by projecting the metric perturbations onto a detector-adapted coordinate system as~\cite{Babak:2006uv}
\bqn
h_{+} &=& (h_{\zeta \zeta} - h_{\iota \iota})/2, \lb{waveform-p}\\
h_{\times} &=& h_{\iota \zeta}.\lb{waveform-c}
\eqn
The components $h_{\zeta \zeta}$, $h_{\iota \iota}$,  and $h_{\iota \zeta}$ are 
\bqn
h_{\zeta \zeta} &=& h_{xx} \cos^2 \zeta - h_{xy}  \sin 2 \zeta + h_{yy} \sin^2 \zeta,\\
h_{\iota \iota} &=&  \cos^2 \iota [h_{xx} \sin^2 \zeta + h_{xy} \sin 2 \zeta + h_{yy} \cos^2 \zeta] + h_{zz} \sin^2 \iota - \sin 2 \iota [h_{xz} \sin \zeta + h_{yz} \cos \zeta], \\
h_{\iota \zeta} &=& \cos \iota \left[ 
 \f{1}{2} h_{xx} \sin 2 \zeta + h_{xy} \cos 2 \zeta - \f{1}{2} h_{yy} \sin 2 \zeta \right] + \sin \iota [h_{yz} \sin \zeta - h_{xz} \cos \zeta], 
\eqn 
where $\iota$ is the inclination angle of the smaller object's orbital plane and $\zeta$ is the longitude of the pericenter. This method provides a good balance between accuracy and computational cost. It captures the key physical features of the waveform and allows us to compare the signals from general relativity and quantum-corrected black hole models.

\begin{figure}[!t]
	\centering
	{\includegraphics[scale =0.23]{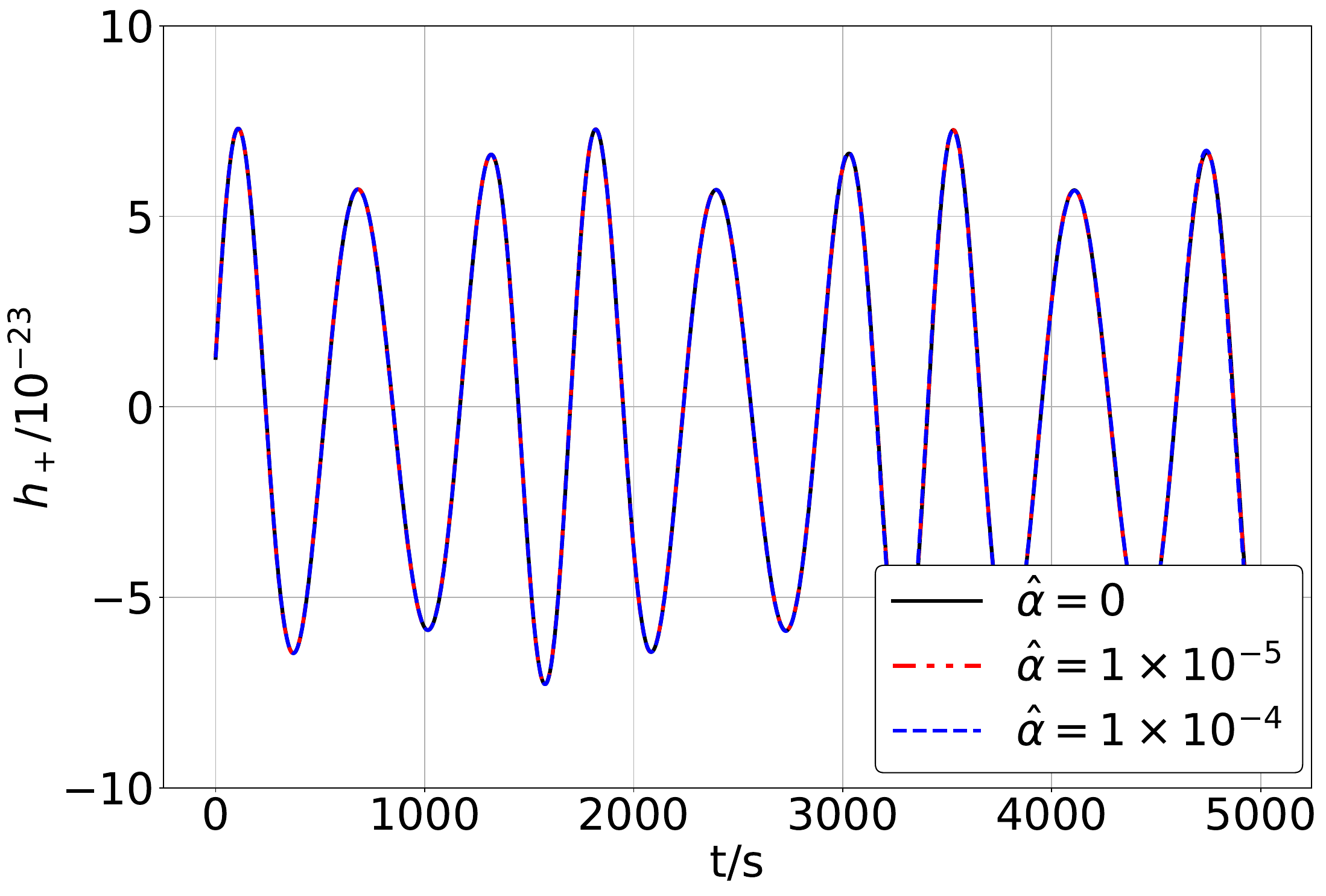}\lb{}}
    {\includegraphics[scale =0.23]{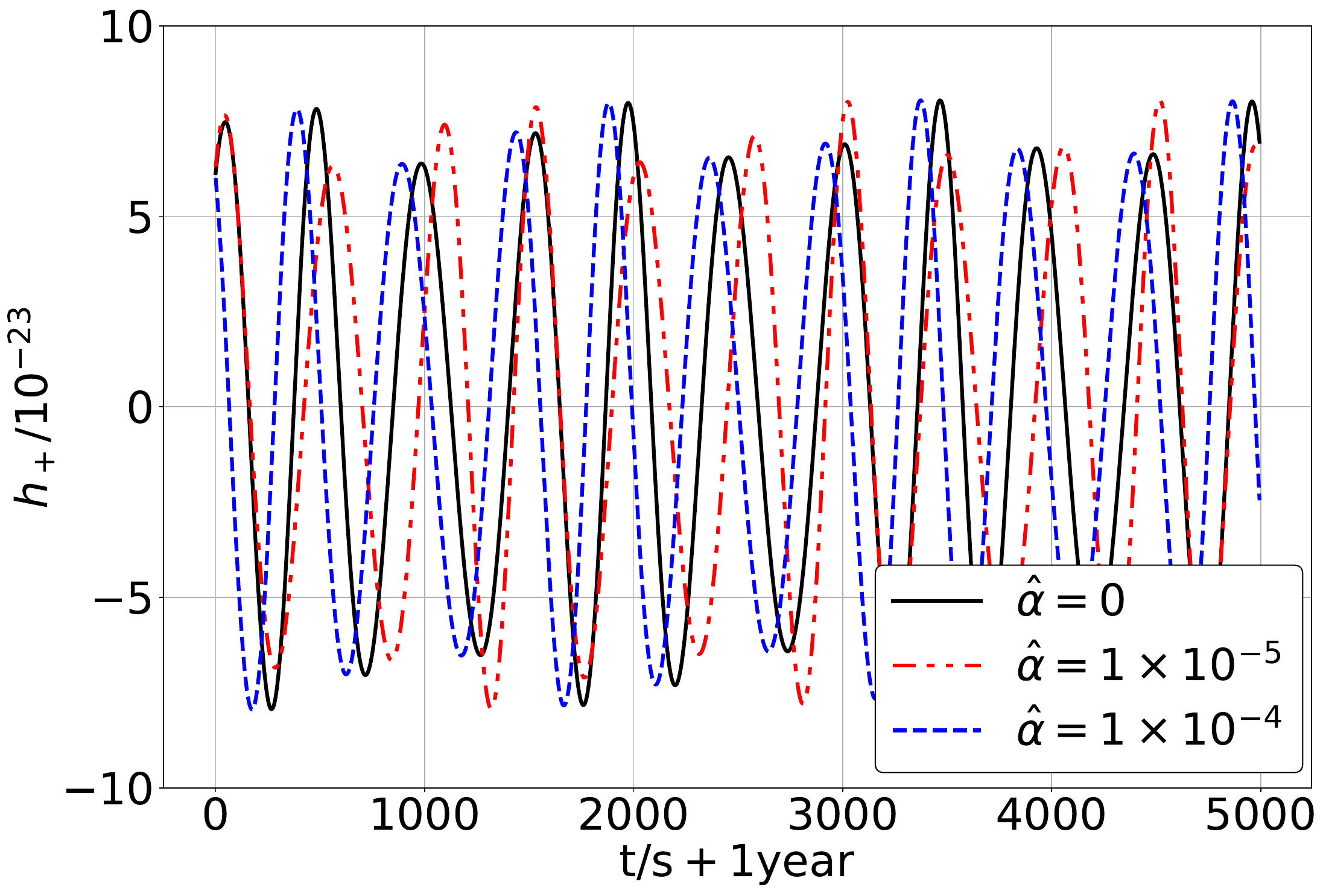}\lb{}}
    {\includegraphics[scale =0.23]{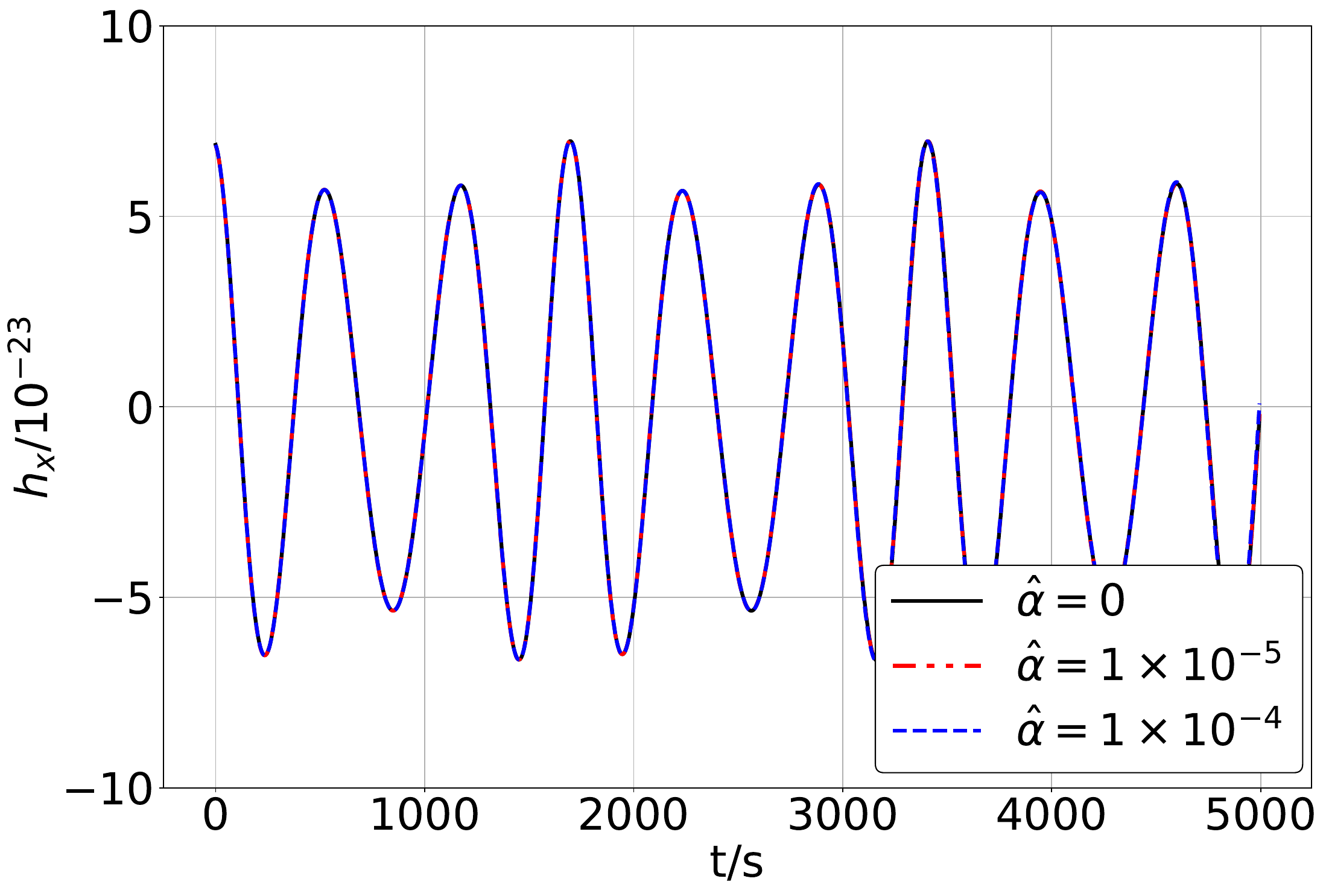}\lb{}}
    {\includegraphics[scale =0.23]{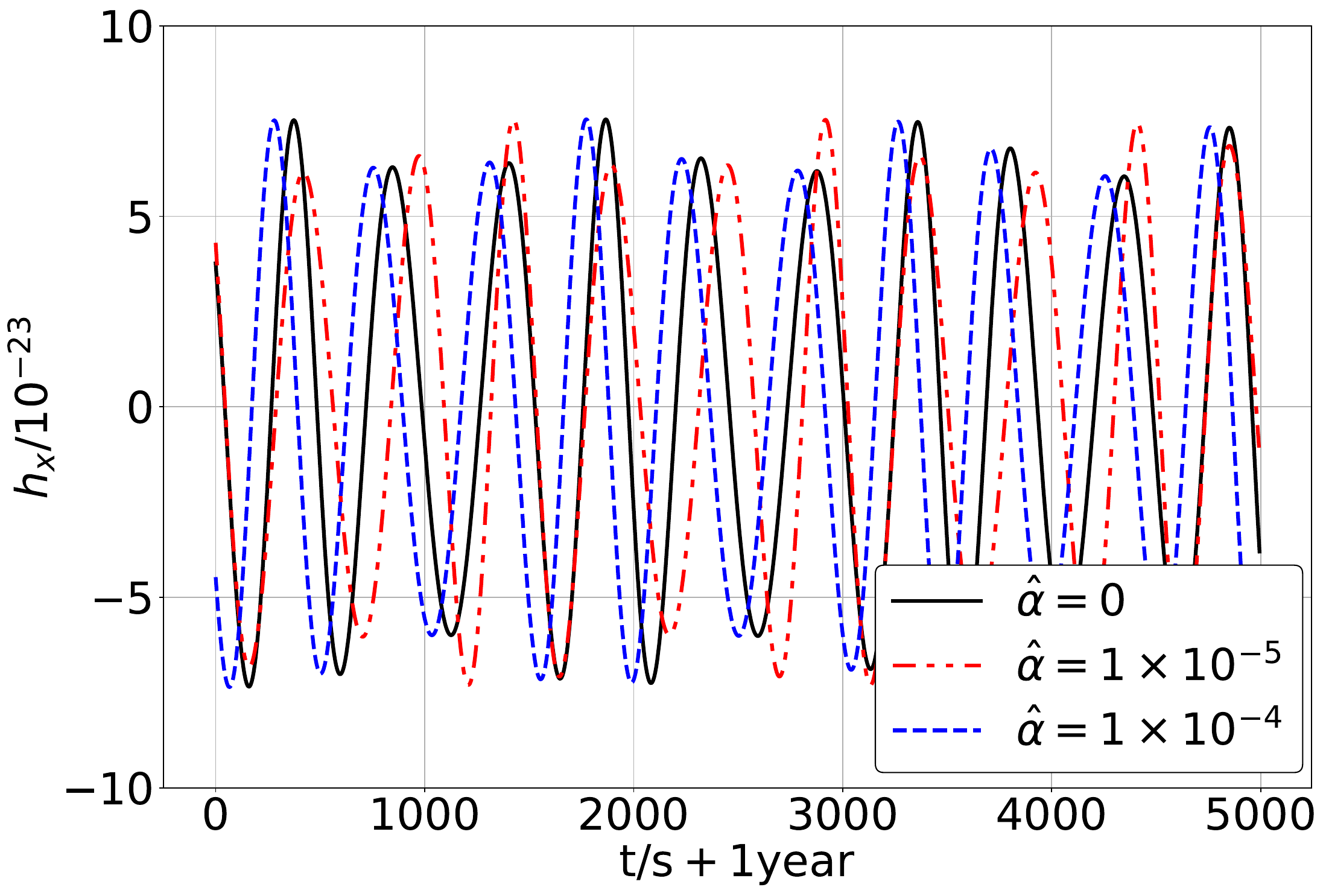}\lb{}}
\caption{Gravitational waveforms from a test object with $m = 10 M_\odot$ along the eccentric orbit with initial $r_p = 10 M$ and $e=0.1$ around a supermassive quantum Oppenheimer-Snyder black hole with $M = 10^6 M_\odot$.}
\label{gravitational-waveforms-1}
\end{figure}

\begin{figure}[!t]
	\centering
	{\includegraphics[scale =0.23]{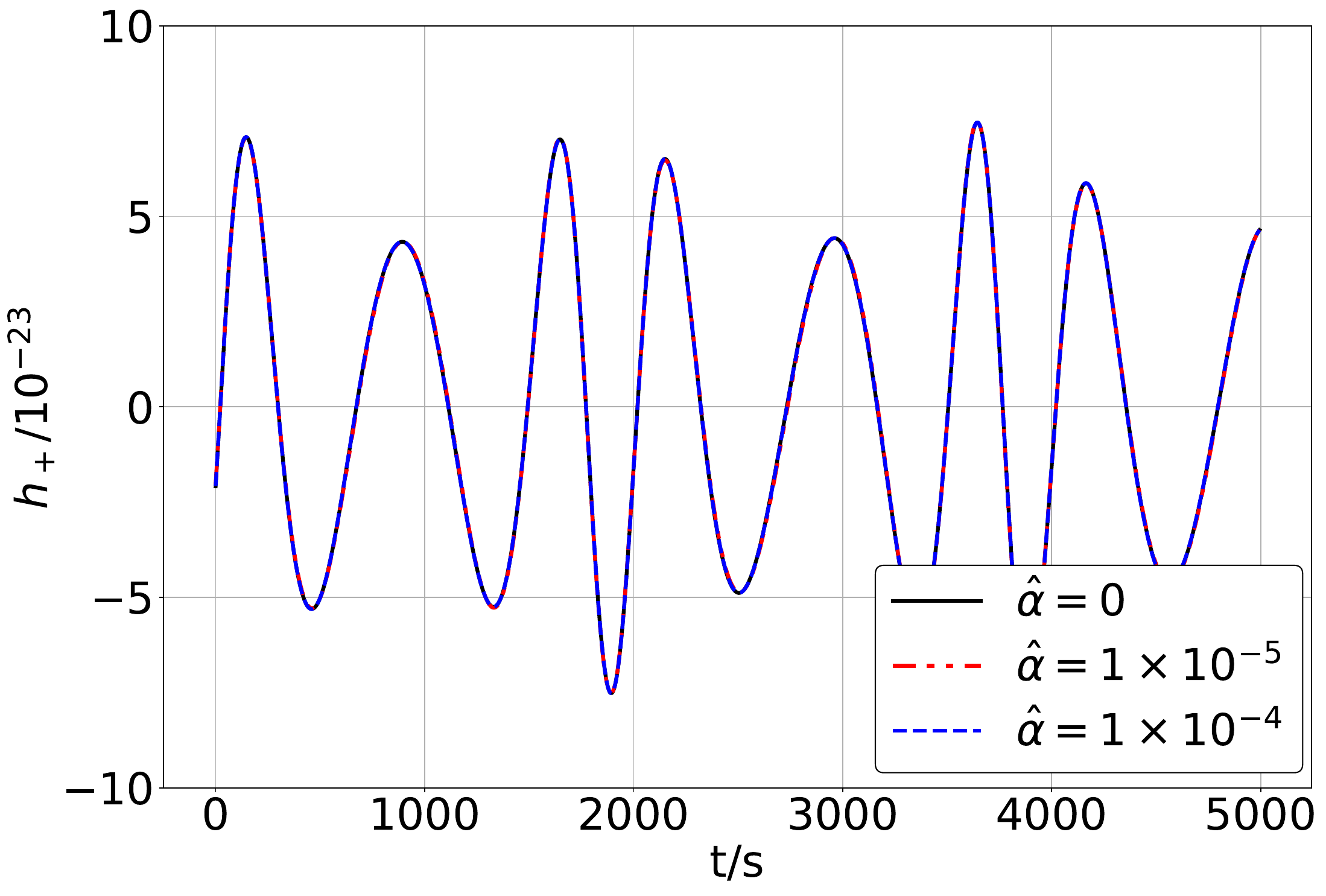}\lb{}}
    {\includegraphics[scale =0.23]{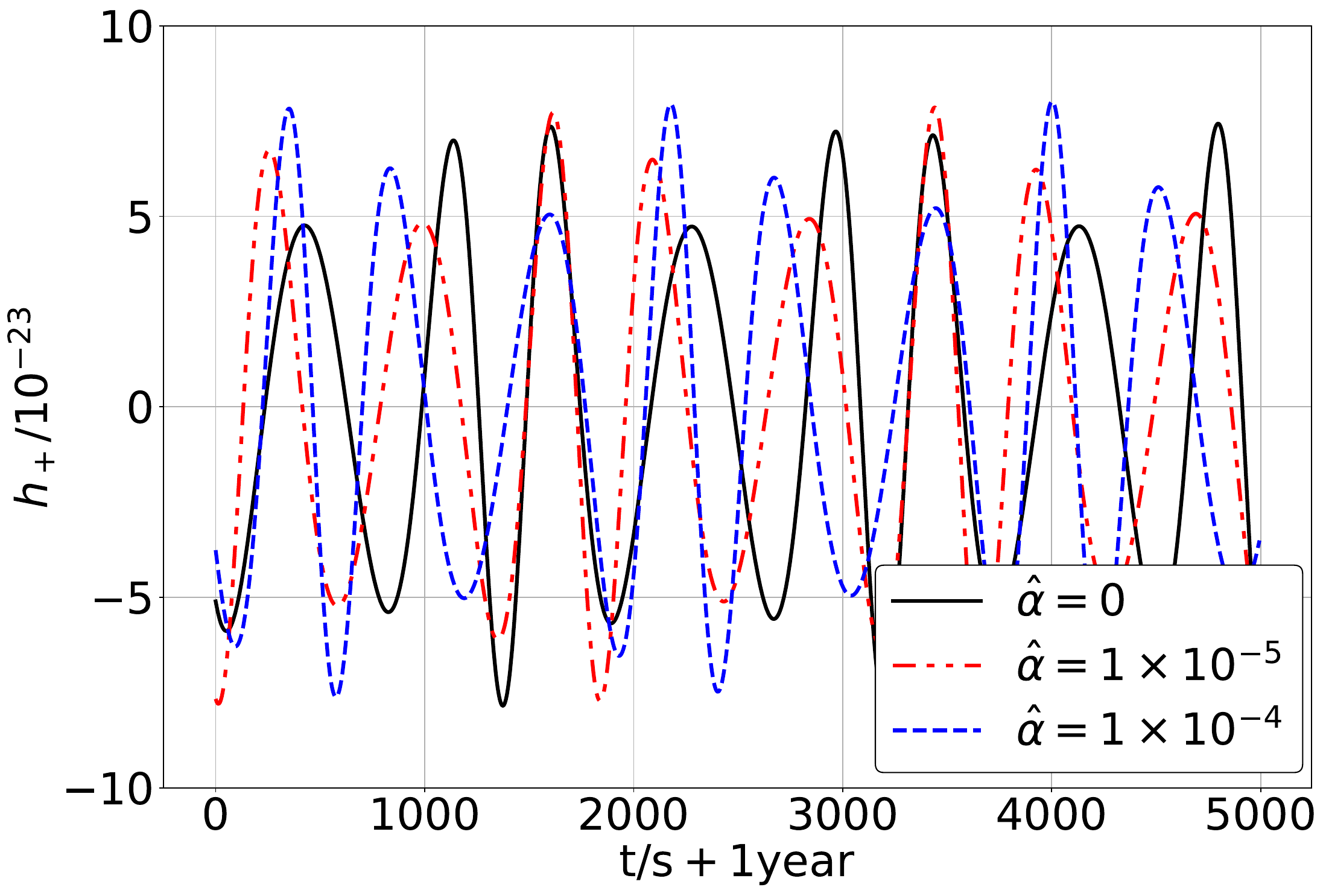}\lb{}}
    {\includegraphics[scale =0.23]{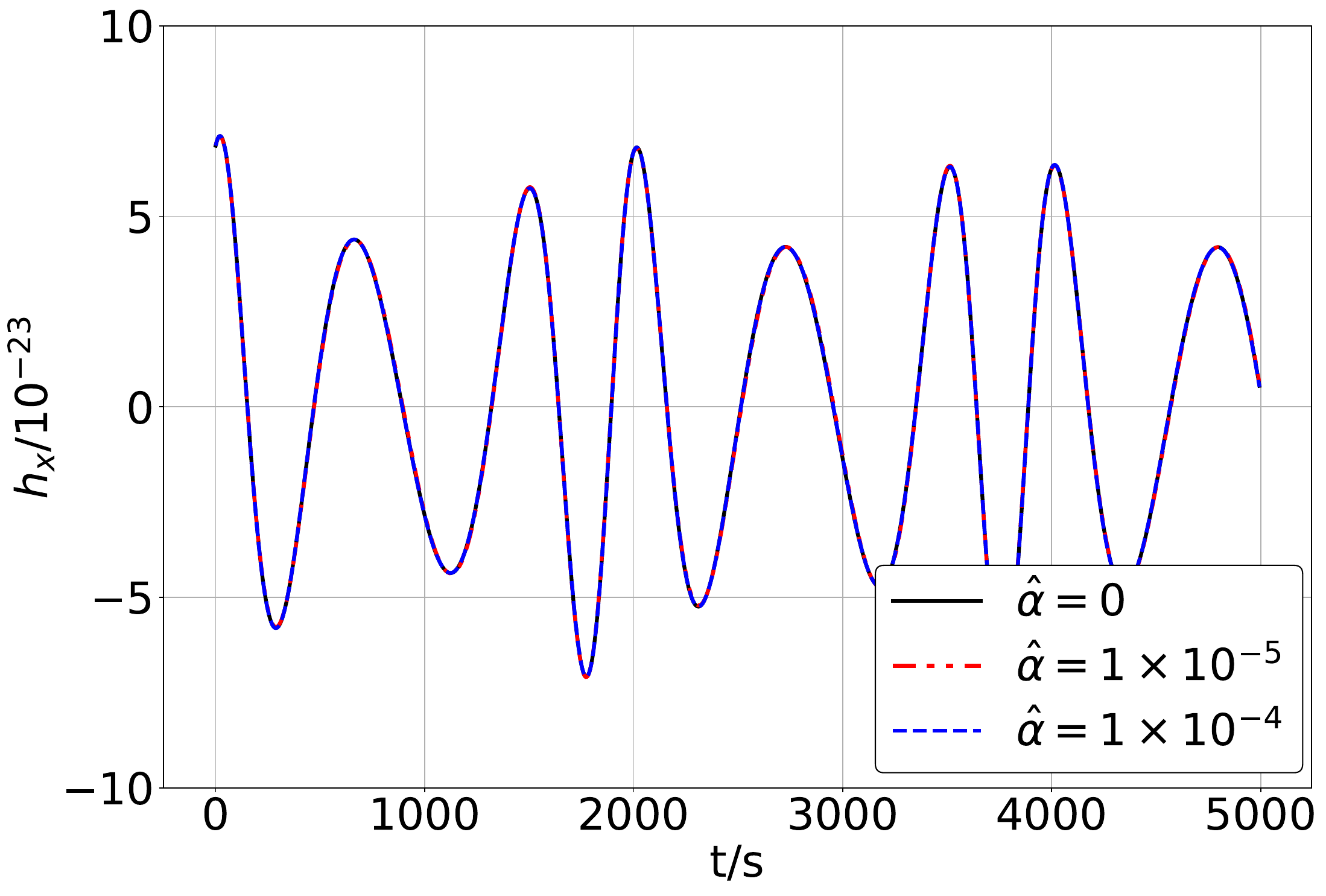}\lb{}}
    {\includegraphics[scale =0.23]{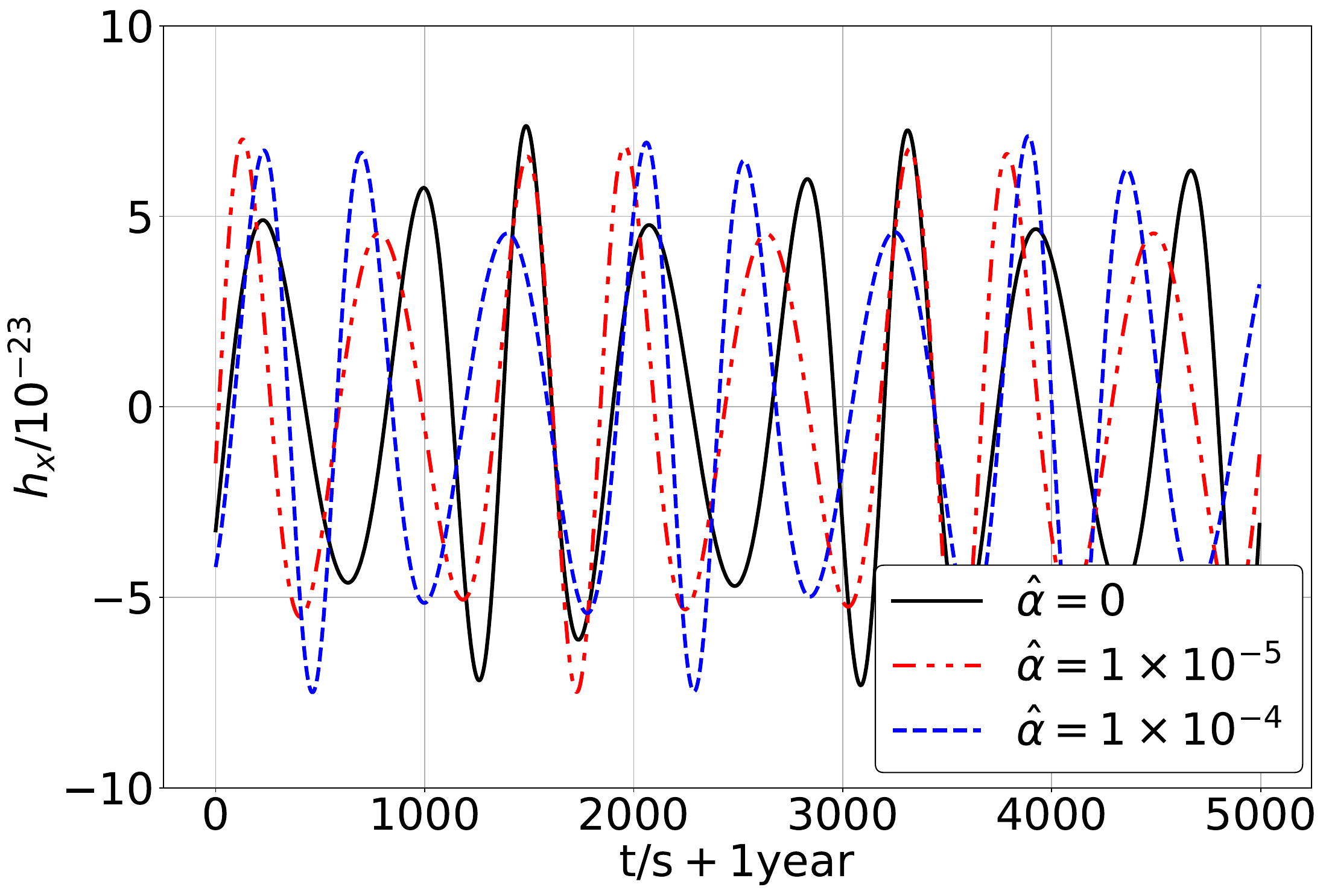}\lb{}}
\caption{Gravitational waveforms from a test object with $m = 10 M_\odot$ along the eccentric orbit with initial $r_p = 10 M$ and $e=0.2$ around a supermassive quantum Oppenheimer-Snyder black hole with $M = 10^6 M_\odot$.}
\label{gravitational-waveforms-2}
\end{figure}

We consider a supermassive quantum-corrected black hole with mass $M=10^6 M_\odot$ and a small object with mass $m=10 M_\odot$. The initial orbital parameters and the initial position of the small object are described in Section 2. We set $\hat{\alpha} = 1\times 10^{-4},~1\times 10^{-5}$ and obtain the orbital trajectories through the numerical algorithm described in Section 2. The external parameters of the extreme mass-ratio system are set as follows: the luminosity distance is $D_\text{L} = 2 \text{Gpc}$, the inclination angle is $\iota = \pi/4$, and the longitude of the pericenter is $\zeta = \pi/4$. Using the orbital trajectories data, we compute the time-domain gravitational waveforms with Eqs.~\eqref{waveform-p} and~\eqref{waveform-c}. The results are plotted in Figs.~\ref{gravitational-waveforms-1} and~\ref{gravitational-waveforms-2}.

The left columns of Figs.~\ref{gravitational-waveforms-1} and~\ref{gravitational-waveforms-2} show the early gravitational waveforms segment (first 5000 seconds), while the right columns present a segment of the waveforms after one year of orbital evolution. One can find that the gravitational waveforms in the background of a Schwarzschild black hole ($\hat{\alpha} = 0$) and those in the background of a quantum-corrected black hole almost completely overlap in the early stage. It indicates that quantum corrections have little impact on short timescales. However, in the late-time panels, noticeable phase differences begin to emerge, especially for larger values of $\hat{\alpha}$. This phase shift becomes more significant as the EMRI system evolves and accumulates the effects of quantum corrections. These results suggest that although quantum gravity effects may be small, they can accumulate over time and lead to observable deviations in the gravitational waveforms of EMRIs.

\section{Spectral analysis and mismatch}\lb{sec4}
\renewcommand{\theequation}{4.\arabic{equation}}
\setcounter{equation}{0} 

In space-based gravitational wave detection, the detector is not stationary but moves around the Sun or Earth. This motion introduces a Doppler modulation to the observed gravitational wave signal over time. For long-lived signals such as those from EMRIs, the Doppler effect can significantly affect the waveform observed by the detector and must be taken into account in data analysis~\cite{Barack:2003fp}. Considering LISA, the detector follows a nearly circular orbit of radius $R \simeq 1 \text{AU}$ with orbital period $T = 1$ year~\cite{LISA:2022yao}.  The Doppler-corrected phase $\Phi_{\text{D}} (t)$ measured by the moving detector is given by~\cite{Barack:2003fp}
\bqn
\Phi_{\text{D}}(t) = \Phi(t) + \Phi(t)' R_\tx{} \sin \theta_\tx{S} \cos\left( 2 \pi t/T - \phi_\tx{S} \right),
\eqn 
where $(\theta,~\phi)$ is the sky location of the source in the detector coordinate system. In this work, we adopt a fixed sky location with $\theta_\tx{S}=\pi/4$ and $\phi_\tx{S}=\pi/4$ for illustration. This phase shift is applied to the waveform computed in Section~\ref{sec3} by replacing $ h(\phi) \rightarrow h(\phi_\text{D})$.

The response of a space-based detector such as LISA to the waveform signals depends on the sky location of the source and the orbital angular momentum direction~\cite{Barack:2003fp}. For a given source, the measured strain in LISA can be written as
\bqn\lb{strain}
h_{\tx{I,II}} = \f{\sqrt{3}}{2}  \left( F^{+}_{\tx{I,II}} h_{+}+ F^{\times}_{\tx{I,II}} h_{\times} \right),
\eqn 
where $F^{+}_{\tx{I,II}}$and $F^{\times}_{\tx{I,II}}$ are the time-dependent antenna pattern functions of the detector for the two gravitational wave polarizations, and the factor $\sqrt{3}/2$ arises from LISA’s equilateral triangular configuration. The explicit expressions of the response function $F^{+}_{\tx{I,II}}$and $F^{\times}_{\tx{I,II}}$ can be found in Ref.~\cite{Barack:2003fp}. The final response signal is $h(t) = h_{\tx{I}} - i h_{\tx{II}}$. In this work, we assume the orbital angular momentum direction of the EMRI systems is $(\theta_{L} = 0,~\phi_{L} = 0)$. Under these assumptions, we calculate the detector response signals of the Doppler-corrected gravitational waveforms with different values of the parameter $\hat{\alpha}$ by applying the full time-dependent modulation from the LISA constellation’s orbit.

To evaluate the detectability of gravitational waves from EMRIs, it is essential to study the waveforms in the frequency domain. We perform a Fast Fourier Transform on the Doppler-corrected gravitational waveforms to obtain the corresponding signals in the frequency domain. For a given source,  the sky and polarization averaged dimensionless characteristic strain is~\cite{Robson:2018ifk}
\bqn
h_\tx{c}(f) = 2 f \left(|\tilde{h}_{+}(f)|^2 + |\tilde{h}_{\times}(f)|^2 \right)^{1/2},
\eqn
where $\tilde{h}_{+,\times} (f)$ is the Fourier transform of $h_{+,\times} (t)$. The characteristic sensitivity for LISA is $(f \text{S}_n(f))^{1/2}$, where $\text{S}_n(f)$ is the  strain spectral sensitivity. We calculate the dimensionless characteristic strain of the gravitational waveforms in Figs.~\ref{gravitational-waveforms-1} and~\ref{gravitational-waveforms-2}.

Moreover, we can compute the inner product between two waveforms in the frequency domain. The noise-weighted inner product between two signals $h_1 (t)$ and $h_2 (t)$ is
\bqn
<h_1|h_2> = 2 \sum_{\lambda = \text{I,II}}\int^{f_\tx{max}} _{f_\tx{min}} \f{\tilde{h}_{1,\lambda}(f) \tilde{h}^\ast_{2,\lambda}(f) + \tilde{h}^\ast_{1,\lambda}(f) \tilde{h}_{2,\lambda}(f) }{S_{n}(f)} df,
\eqn
where $\tilde{h}_i (f)$ is the Fourier transform of the strain signal, and 
$S_n (f)$ is the one-sided power spectral density of the detector noise. In this work, we use the official LISA sensitivity curve for $S_n (f)$ in Ref.~\cite{Robson:2018ifk}, and set the integration range to cover the detector’s sensitive frequency band. This inner product forms the basis for calculating both the SNR and the waveform overlap. The optimal SNR for a signal $h$ is given by $\rho= 
\sqrt{⟨h|h⟩}$. A higher SNR implies a stronger signal relative to the detector noise and generally corresponds to better parameter estimation accuracy.

To quantify how similar two gravitational waveforms are, we use the overlap, which measures the normalized inner product between two signals in the frequency domain. Given two time-domain detector response signals $h_1(t)$ and 
$h_2(t)$, we compute their Fourier transforms 
$\tilde{h}_1 (f)$ and 
$\tilde{h}_2 (f)$, and define the overlap as
\bqn\lb{overlap}
\mathcal{O}(h_1, h_2) =  \frac{<h_1|h_2>}{\sqrt{<h_1|h_1><h_2|h_2>}}.
\eqn
However, in realistic scenarios, two gravitational waveforms may differ by a constant phase shift or a time translation due to different choices of initial conditions. To remove these unphysical differences, we compute the maximized overlap $\mathcal{O}_{\tx{max}}$ by maximizing the overlap over a constant time shift 
$\Delta t$ and phase offset $\Delta \phi$:
\bqn\lb{faithfulness function}
\mathcal{O}_{\tx{max}}(h_1, h_2) = \mathop{\text{max}}\limits_{\{\Delta t,\Delta \phi \}} \frac{<h_1|h_2>}{\sqrt{<h_1|h_1><h_2|h_2>}}.
\eqn
This maximized overlap ensures that the comparison captures only intrinsic differences in waveform shape and phase evolution, rather than arbitrary offsets in time or phase. Using the maximized overlap, we define the mismatch as
\bqn
\mathcal{M}(h_1, h_2) = 1 - \mathcal{O}_{\tx{max}}(h_1, h_2).
\eqn
The mismatch gives a quantitative measure of distinguishability between waveforms. When the mismatch value is close to zero, the waveforms can be regarded as nearly identical; conversely, a larger mismatch reflects pronounced differences. The threshold for mismatch is $\mathcal{M}_c = N/(2 \rho^2)$, where $N$ is the dimension of the intrinsic parameters of the gravitational waveform model and $\rho$ is the SNR of the gravitational wave signals being compared. Given the dimension of the model parameters and the assumed SNR of the gravitational wave signal, two gravitational waveforms are observationally indistinguishable if their mismatch is below the threshold.

The intrinsic parameters of the gravitational waveform model in this work are: the initial orbital pericenter distance $r_{p0}$, the initial orbital eccentricity $e_0$, the supermassive black hole's mass $M$, the small object’s mass $m$, the initial azimuthal angle $\phi_0$, and the quantum correction parameter $\hat{\alpha}$. So the dimension of the model parameters is $N = 6$. Here, we assume the SNR of the gravitational wave signal is $\rho = 10$. Then the threshold of mismatch in our case is $\mathcal{M}_c = 0.03$. We calculate the gravitational waveforms in the background of a Schwarzschild black hole (corresponding to $\hat{\alpha}=0$) and the waveforms with different values of the quantum correction parameter $\hat{\alpha}$, rescale the luminosity distance of these waveforms with $\rho = 10$, and compute the mismatch between these waveforms' response signals. For each case, we use the waveforms from the final year of inspiral. The mismatch is computed in the LISA frequency band for each waveform pair.

\begin{figure}[!h]
	\centering
    {\includegraphics[scale =0.24]{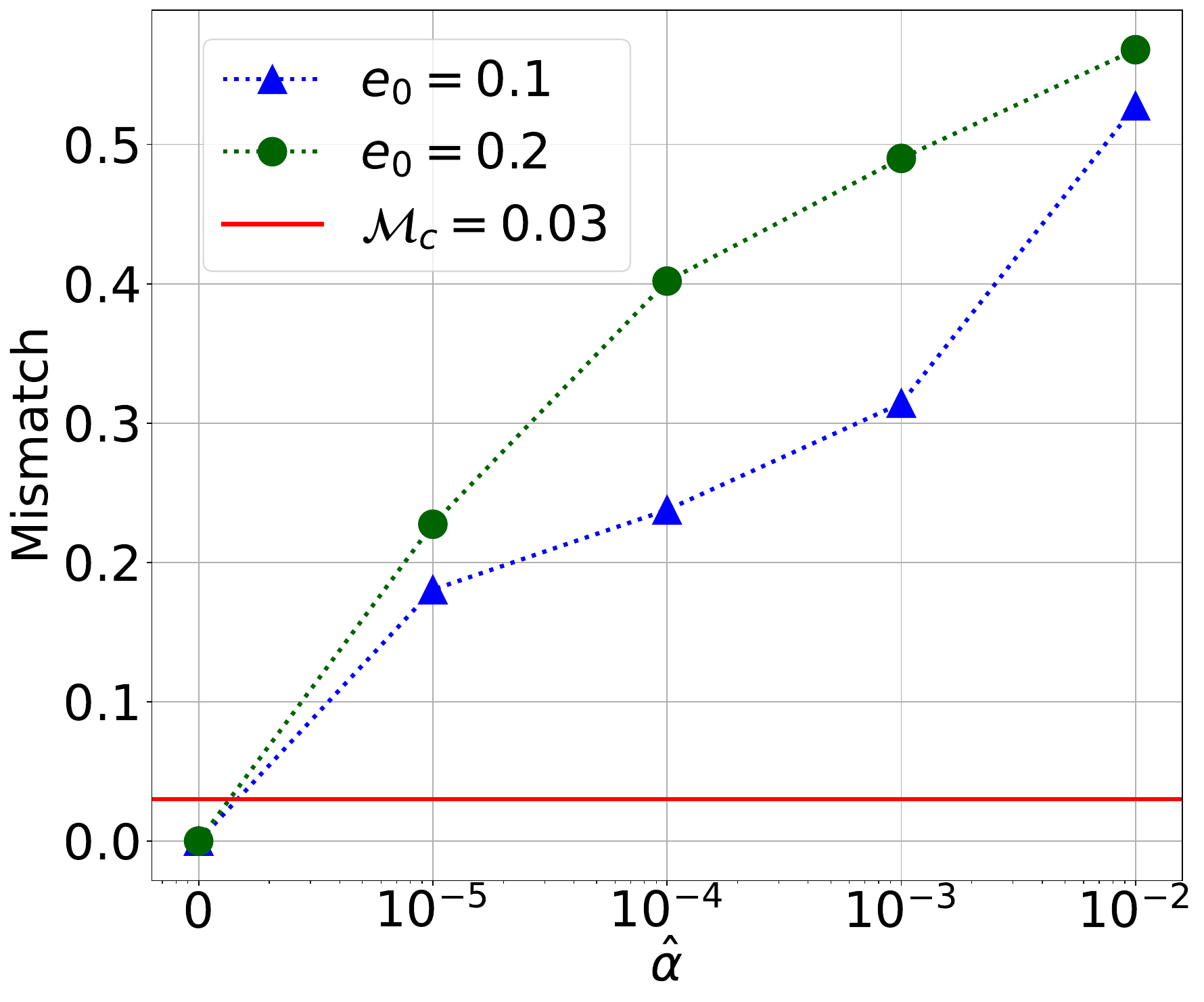}}\lb{}
	\caption{Mismatch between the final-year gravitational wave signals from eccentric EMRIs in the backgrounds of a quantum-corrected black hole and a Schwarzschild black hole.}
	\label{plot-mismatch}
\end{figure}

Due to the limitations in the accuracy of our calculations for the orbital evolution and corresponding gravitational waveforms of the EMRI system, we set the nonzero value of the quantum correction parameter to $1 \times 10^{-5}$. We plot the mismatch between the gravitational wave signals with $\hat{\alpha} = 0$ and the signals with different values of $\hat{\alpha}$ in Fig.~\ref{plot-mismatch}. Our results show that as $\hat{\alpha}$ increases, the mismatch also increases. This is mainly due to cumulative dephasing effects introduced by quantum corrections to the background spacetime. Even when the parameter $\hat{\alpha}$ is set to a very small value of $1 \times 10^{-5}$, the mismatch still exceeds the threshold. This indicates that future space-based gravitational wave observations of EMRIs could constrain the parameter $\hat{\alpha}$ to at least the order of $O(10^{-5})$. This is significantly more stringent than the constraint $\hat{\alpha} <1.4087$ given by black hole shadow observations~\cite{Zhao:2024elr}.

\begin{figure}[!h]
	\centering
		\includegraphics[scale =0.24]{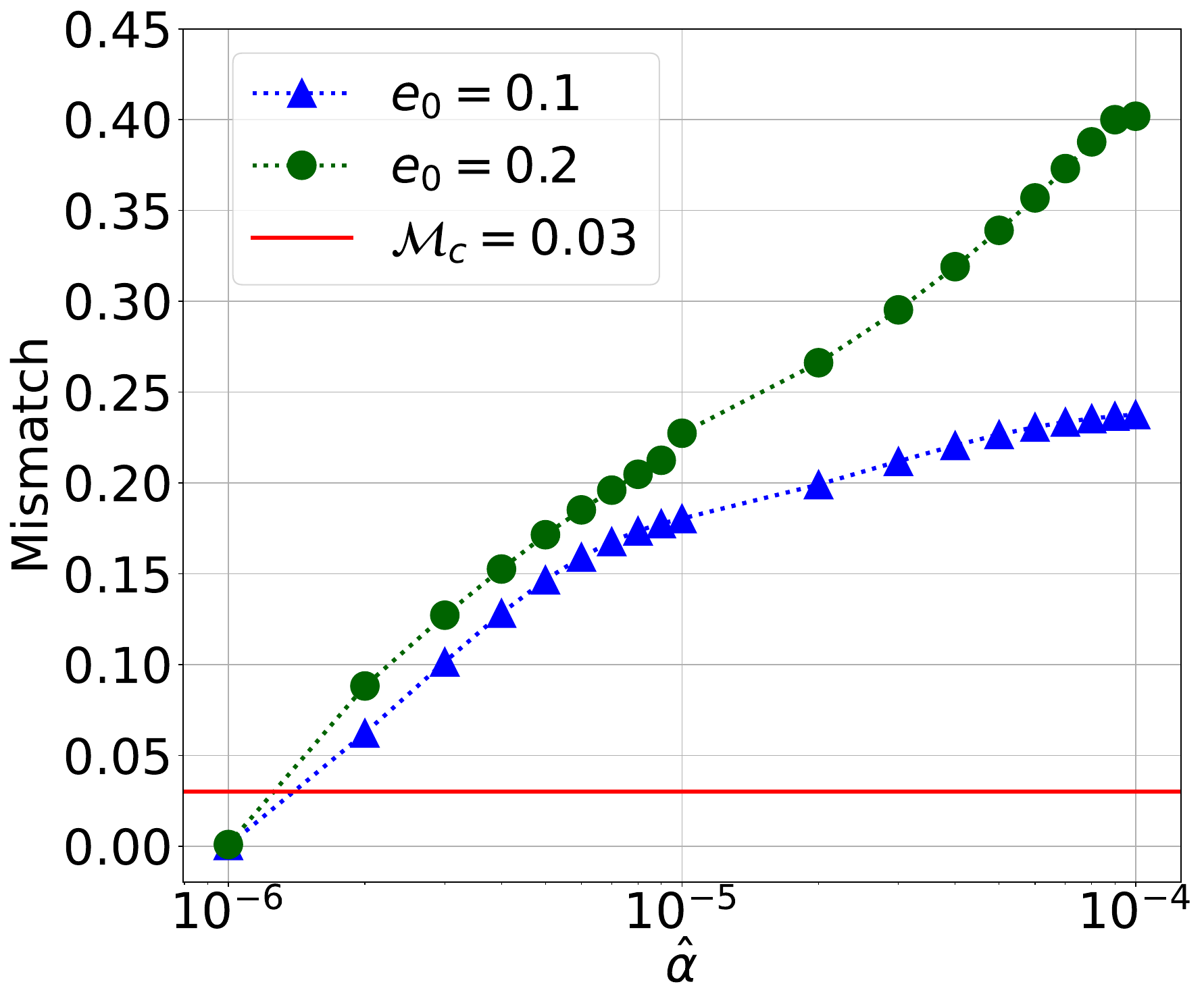}
		\caption{Mismatch as a continuous function of the quantum-correction parameter $\hat{\alpha}$ from $10^{-6}$ to $10^{-4}$ for two representative initial eccentricities, $e_0=0.1$ and $e_0=0.2$. }
		\label{plot-newmismatch}
\end{figure}

To quantify the dependence of the waveform mismatch on the quantum-correction parameter $\hat{\alpha}$, we extend our analysis by computing the mismatch for a continuous range of values from $10^{-6}$ to $10^{-4}$. The results are shown in Fig.~\ref{plot-newmismatch}. It provides a clearer view of the scaling behavior across small $\hat{\alpha}$. For two representative initial eccentricities, $e_0=0.1$ and $e_0=0.2$, the mismatch increases monotonically as $\hat{\alpha}$ grows. Notably, the mismatch crosses the detectability threshold $ \mathcal{M}_c = 0.03$ when $\hat{\alpha}$ increases from approximately $1 \times 10^{-6}$ to $2 \times 10^{-6}$. This behavior identifies the parameter range in which the accumulated dephasing between the quantum-corrected and Schwarzschild EMRI waveforms becomes observationally significant. These results strengthen our constraint estimates and clarify the sensitivity of long-duration EMRIs to small quantum-gravity corrections.

\begin{figure}[!t]
	\centering
	\subfigure[$e_0 = 0.1$]{\includegraphics[scale =0.29]{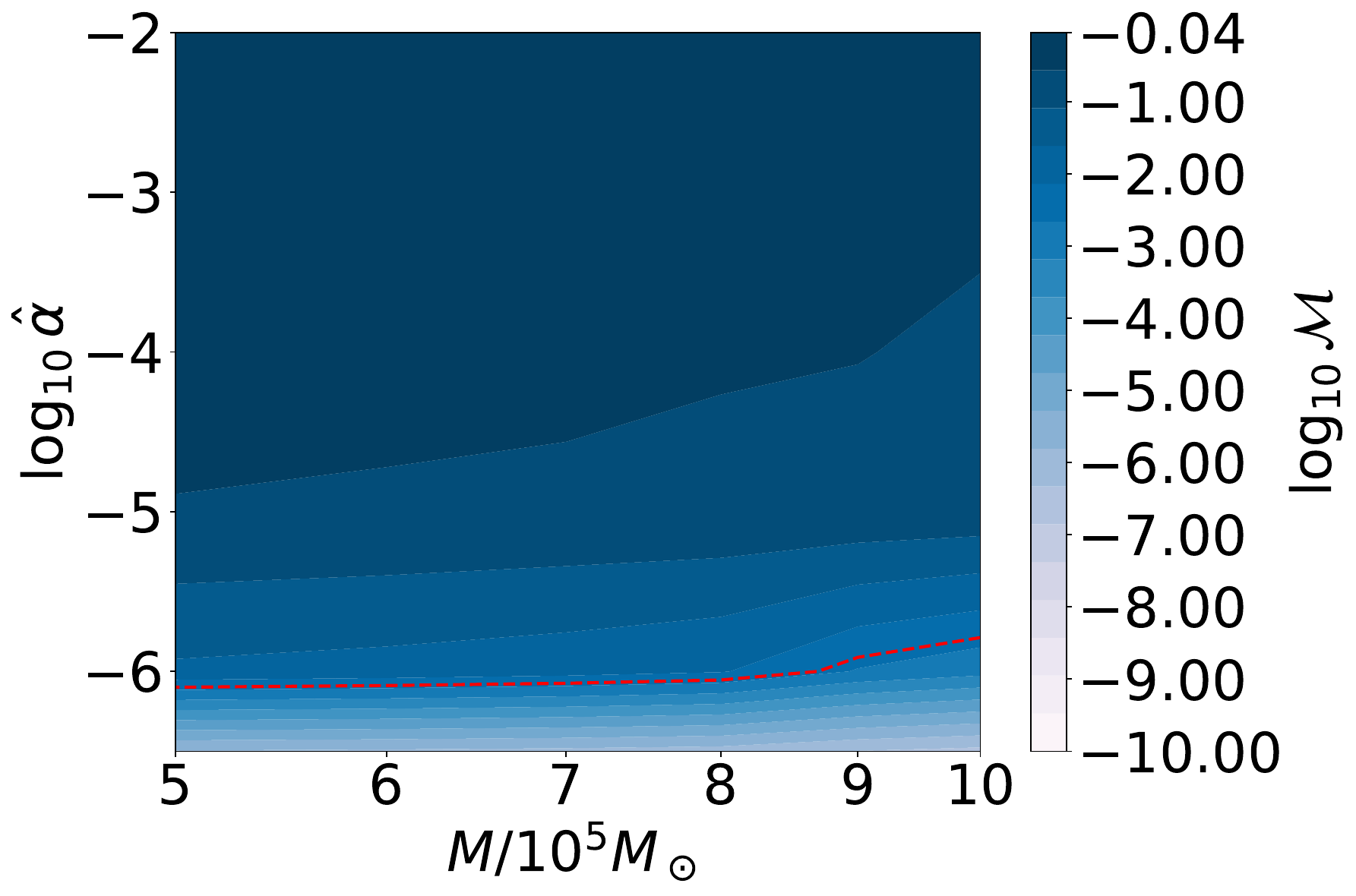}}
	\subfigure[$e_0 = 0.2$]{\includegraphics[scale =0.29]{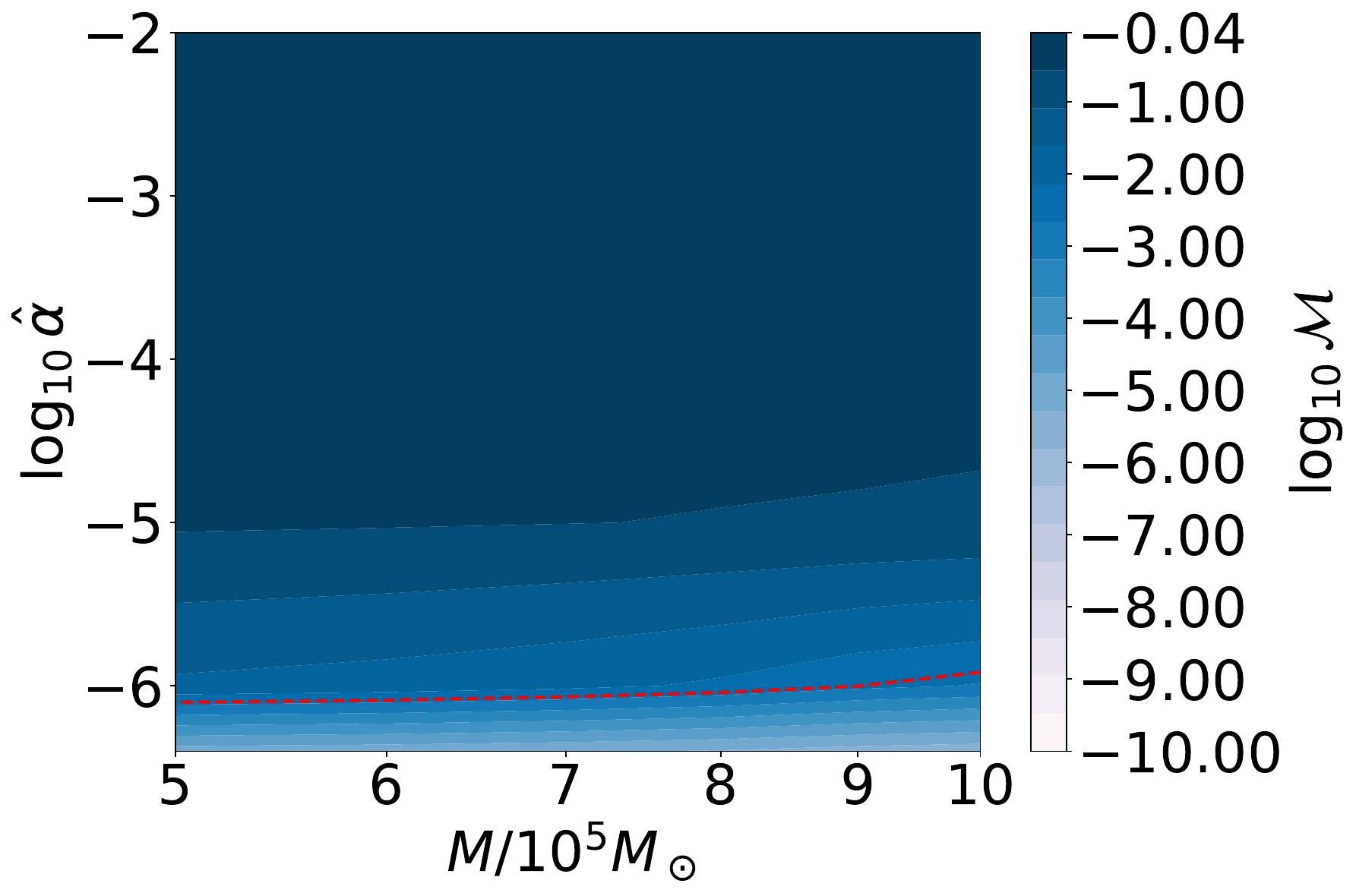}}
	\caption{Mismatch distribution in the $(M,~ \hat{\alpha})$ parameter space. The red curve marks the detectability threshold $ \mathcal{M}_c = 0.03$.}
	\label{contour}
\end{figure}

To further examine possible parameter degeneracies in our model, we analyze the mismatch distribution in the $(M,~ \hat{\alpha})$ space. The results are shown in Fig.~\ref{contour}. A mild degeneracy between the central black hole's mass and the quantum parameter is visible through the slight tilting of the mismatch contours. This indicates that small variations in $M$ can partially compensate for the phase shifts induced by $\hat{\alpha}$. However, the threshold curve remains nearly horizontal across the explored region, showing that the effect of $M$ cannot fully mimic the influence of $\hat{\alpha}$. Therefore, the constraint on $\hat{\alpha}$
obtained in this work remains robust against this limited degeneracy. We also note that, in realistic EMRI observations, the spin of the central black hole may introduce additional degeneracies with $\hat{\alpha}$. Since the present study is based on a static and spherically symmetric quantum-corrected metric, such spin-related effects are not included here. A rotating quantum-corrected spacetime and the associated waveform model are needed to quantify the impact of black hole spin on the parameter estimation. We plan to extend our analysis to the spinning case in future work.

\section{Conclusions and Discussions}\lb{sec5}

In this work, we explored the potential of using gravitational waves from EMRIs in eccentric orbits to constrain the quantum Oppenheimer-Snyder black holes. We began by modeling the orbital evolution of a stellar-mass object around a supermassive black hole under the influence of gravitational radiation. To accurately describe the gradual inspiral, we employed the adiabatic approximation in combination with the mass-quadrupole radiation formula. The orbital energy and angular momentum losses were computed numerically, and the evolution of the semi-latus rectum $p$ and eccentricity $e$ was tracked over time. We found that the semi-latus rectum $p$ decreases
monotonically and at an accelerating rate over time due to gravitational radiation. However, the eccentricity $e$ first decreases gradually
and then increases again during the late stages of evolution. It is indicated that the quantum correction parameter $\hat{\alpha}$ slightly alters the rate of orbital evolution. We also investigated the effect of $\hat{\alpha}$ on the orbital phase. We found that quantum corrections induce cumulative phase shifts, which may lead to observable modifications in the dynamics and gravitational waveforms of EMRIs. Using these evolving trajectories, we then constructed time-domain gravitational waveforms via the numerical kludge method. Our results demonstrated that while quantum corrections have little impact on EMRI waveforms at early times, their cumulative effects may lead to observable deviations at late stages.

To assess the potential of future space-based gravitational wave detectors in constraining the quantum parameter $\hat{\alpha}$, we further analyzed the detector's response to the gravitational waves from EMRIs in the background of the supermassive quantum Oppenheimer-Snyder black holes. We accounted for the Doppler modulation of the waveform phase resulting from the detector's orbital motion. We set the luminosity distance to 2 Gpc, computed the frequency-domain dimensionless characteristic strains. We then fixed the SNR to 10 and rescaled the luminosity distance accordingly to maintain this value across different parameter choices. Based on these conditions, we computed the mismatch between the gravitational wave response signals for different values of the quantum correction parameter $\hat{\alpha}$. We found that the mismatch exceeds the typical detection threshold (e.g., $0.03$) even for small values such as $\hat{\alpha} = 1 \times 10^{-5}$. These findings suggest that future space-based detectors like LISA will be capable of distinguishing the subtle phase differences induced by quantum corrections and could impose much stronger constraints on the parameter $\hat{\alpha}$ than those derived from black hole shadow observations or perihelion precession measurements. This highlights the unique power of EMRIs as a probe for quantum gravity and a window into the fundamental nature of black hole spacetimes.

This work provides insights into the effects of quantum corrections on EMRIs, but several aspects warrant further exploration. One limitation is that we only considered the motion of the smaller object in the equatorial plane. In reality, the orbits of smaller objects are typically inclined~\cite{Amaro-Seoane:2007osp, Amaro-Seoane:2012aqc, Qiao:2024gfb}. Additionally, the black hole studied in this work is spherically symmetric, while real astrophysical black holes are rotating, which would cause a range of interesting effects~\cite{Barack:2006pq, Glampedakis:2002ya}. Moreover, the astrophysical environments surrounding supermassive black holes affect the smaller objects' orbital dynamics~\cite{Rahman:2023sof, Zi:2025lio}. EMRIs around supermassive black holes with accretion disks may also produce electromagnetic radiations, such as X-ray emissions~\cite{Sukova:2021thm, Xian:2021xkc}. Combining gravitational wave detections with electromagnetic observations could offer rich science opportunities to study the black hole physics and cosmology~\cite{Pan:2021oob, Lyu:2024gnk}. We will address these issues in our future work.

\section*{Acknowledgements}

We thank Chao Zhang and Guoyang Fu for the important discussions. This work was supported by the National Key Research and Development Program of China (Grant No.~2021YFC2203003), the National Natural Science Foundation of China (Grants No.~12475056, No.~12575055, No.~12275238, and No.~12247101), the National Natural Science Foundation of Gansu Province (Grant No.~22JR5RA389), the `111 Center' (Grant No.~B20063), Talent Scientific Fund of Lanzhou University, and Gansu Province's Top Leading Talent Support Plan.

\appendix

\textcolor{black}{
\section{Motion of the object on the equatorial plane}\lb{AppA}}
\renewcommand{\theequation}{A.\arabic{equation}}
\setcounter{equation}{0}

The equations of motion for an object on the equatorial plane around the supermassive quantum Oppenheimer-Snyder black hole are derived in Ref.~\cite{Yang:2024lmj} as
\bqn
\f{d t}{d \tau} &=& \f{E}{ 1 - \f{2 M}{r} + \f{\alpha M^2}{r^4} }, \lb{dott}\\
\f{d \phi}{d \tau} &=& \f{L}{r^2}, \lb{dotphi} \\
\lf( \f{d r}{d \tau} \rt)^2 &=& E^2 - V_\text{eff}, \lb{dotr}
\eqn 
where $\tau$ is the proper time, and $E$ and $L$ represent the test particle's energy and orbital angular momentum per unit mass, respectively. The radial effective potential in Eq.~\eqref{dotr} is
\bqn
V_\text{eff} = \lf( 1 - \f{2 M}{r} + \f{\alpha M^2}{r^4}  \rt) \lf( 1+\f{L^2}{r^2} \rt).
\eqn

The motion of the test object can be well described by Keplerian orbital parameters~\cite{gravity-book}. The distance from the smaller object to the central supermassive black hole can be rewritten as
\bqn\lb{kepler}
r = \f{p}{1 + e \cos \Psi},
\eqn 
where $p$ is the semi-latus rectum of the orbit of the smaller object, $e$ is the eccentricity, and $\Psi$ is the intersection angle between the semi-major axis and the radial of the orbit. The pericenter and apocenter of the orbit are given by
\bqn
r_{a} = p/(1-e), ~~ r_{p} = p/(1+e).
\eqn
Because the radial velocity of the test object vanishes at both the pericenter and apocenter, for an orbit with fixed energy $E$ and orbital angular momentum $L$, one can get the pericenter $r_p$ and apocenter $r_a$ of the orbit from 
\bqn
E^2 - \lf( 1 - \f{2 M}{r} + \f{\alpha M^2}{r^4}  \rt) \lf( 1+\f{L^2}{r^2} \rt) = 0.
\eqn 
Then, the semi-latus rectum of the orbit is
\bqn\lb{p}
p = \frac{2 r_a r_p}{r_a + r_p},
\eqn
and the eccentricity of the orbit is  
\bqn\lb{e}
e = \frac{r_a - r_p}{r_a + r_p}.
\eqn

With the condition that the radial velocity of the smaller object vanishes at both the pericenter and the apocenter, one can get the energy $E$ and orbital angular momentum $L$ from Eq.~\eqref{dotr} as
\bqn
E^2 &=& \f{f(r_a) f(r_p) (r_a^2 - r_p^2)}{f(r_p) r_a^2 - f(r_a) r_p^2}, \lb{E-eq} \\
L^2 &=& \f{r_a^2 r_p^2 [f(r_a) - f(r_p)]}{f(r_p) r_a^2 - f(r_a) r_p^2}. \lb{L-eq}
\eqn 
In each period of the orbit, the apsidal angle $\Delta \phi$ passed by the smaller object is
\bqn\lb{apsidal-angle}
\Delta \phi = 2 \int_{r_p}^{r_a} \f{d \phi}{dr} dr = 2 \int_{0}^{\pi} \f{d \phi}{d \Psi} d \Psi,
\eqn 
where the integral kernel can be obtained with Eqs.~\eqref{dotphi}, \eqref{dotr} and \eqref{kepler} as 
\bqn\lb{dphidpsi-1}
\f{d \phi}{d \Psi} = \f{d \phi / dr}{d \Psi /dr} =  \f{ e p L \sin \Psi}{r^2 (1+e \cos \Psi)^2 \sqrt{E^2 - V_\text{eff}}}.
\eqn 
Taking the energy $E$ \eqref{E-eq} and orbital angular momentum $L$ \eqref{L-eq} into Eq.~\eqref{dphidpsi-1}, and expanding the result to the leading order of the parameter $\hat{\alpha}$, one can get
\bqn\lb{dphidpsi-2}
\f{d \phi}{d \Psi} &=& 1 + \f{(3+e \cos \Psi) M}{p} + \f{3 M^2 (3 + e \cos \Psi)^2}{2 p^2} \nb\\
&&+ \f{135 M^3 - (6 + e^2) M^3 \hat{\alpha} + e M^3 (135 - 4 \hat{\alpha}) \cos \Psi + e^2 M^3 (45 - \hat{\alpha}) \cos^2 \Psi + 5 e^3 M^3 \cos^3 \Psi}{2 p^3}.
\eqn 
Then, integrating Eq.~\eqref{apsidal-angle} with Eq.~\eqref{dphidpsi-2}, one can obtain the perihelion precession of the smaller object's orbit around the supermassive quantum Oppenheimer-Snyder black hole in each period as
\bqn\lb{precession}
\Delta \omega_{\tx{quantum}} = \Delta \phi - 2 \pi = \f{6 \pi M}{p} + \f{3 (18 + e^2) M^2 \pi}{2 p^2} + \f{3 \pi M^3 (90 + 15 e^2 - e^2 \hat{\alpha} - 4 \hat{\alpha})}{2 p^3}.
\eqn 
The first and second terms on the right side of the second equality in Eq.~\eqref{precession} are just the perihelion precession around a Schwarzschild black hole, and the third term is from the quantum corrections. The ratio of the perihelion precession around the supermassive quantum Oppenheimer-Snyder black hole to that of the Schwarzschild black hole is
\bqn
f_{\tx{sp}} = \f{\Delta \omega_{\tx{quantum}}}{\Delta \omega_{\tx{GR}}} = 1 + \f{M(18+e^2)}{4p} + \f{M^2 (90 + 15 e^2 - e^2 \hat{\alpha} - 4 \hat{\alpha})}{4 p^2}.
\eqn
GRAVITY Collaboration has observed the complete orbits of the S2 star and found that the ratio of the measured perihelion precession of the S2 star's orbit to the one predicted by general relativity as \cite{GRAVITY:2020gka}
\bqn\lb{sp1}
f_{\tx{sp}} = \f{\Delta \omega_{\tx{s2}}}{\Delta \omega_{\tx{GR}}} = 1.10 \pm 0.19.
\eqn
Some properties of the system of S2 star's orbit are: the mass of SgrA$^\ast$ $M_{\tx{S}} = 4.3 \times 10^{6}~M_\odot$,  the luminosity distance $D = 8.35~\tx{kpc}$, the eccentricity $e = 0.884649$, and the semi-major axis $a = p/(1-e^2) = 125.058~\tx{mas}$. With these observational results, we can find that the upper bound on the parameter $\hat{\alpha}$ from Eq.~\eqref{sp1} is of order $\mathcal{O}(10^6)$. It indicates that the orbital perihelion precession cannot provide a meaningful constraint on this parameter.


\begin{thebibliography}{399}

\bibitem{LISA:2017pwj}
P.~Amaro-Seoane \textit{et al.} [LISA],
Laser Interferometer Space Antenna, \href{https://arxiv.org/abs/1702.00786}{arXiv:1702.00786 [astro-ph.IM]}.

\bibitem{Hu:2017mde}
W.~R.~Hu and Y.~L.~Wu,
The Taiji Program in Space for gravitational wave physics and the nature of gravity,
\href{\doibase 10.1093/nsr/nwx116}{Natl. Sci. Rev. \textbf{4}, 5, 685-686 (2017)}.

\bibitem{TianQin:2015yph}
J.~Luo \textit{et al.} [TianQin], TianQin: a space-borne gravitational wave detector, \href{\doibase 10.1088/0264-9381/33/3/035010}{Class. Quant. Grav. \textbf{33}, 3, 035010 (2016)},
arXiv:1512.02076 [astro-ph.IM].

\bibitem{Musha:2017usi}
M.~Musha [DECIGO Working group], Space gravitational wave detector DECIGO/pre-DECIGO, \href{\doibase 10.1117/12.2296050}{Proc. SPIE Int. Soc. Opt. Eng. \textbf{10562}, 105623T (2017)}.

\bibitem{Apostolatos:1993nu}
T.~Apostolatos, D.~Kennefick, E.~Poisson, and A.~Ori, Gravitational radiation from a particle in circular orbit around a black hole. 3: Stability of circular orbits under radiation reaction,
\href{\doibase 10.1103/PhysRevD.47.5376}{Phys. Rev. D \textbf{47}, 5376-5388 (1993)}.

\bibitem{Tanaka:1993pu}
T.~Tanaka, M.~Shibata, M.~Sasaki, H.~Tagoshi, and T.~Nakamura, Gravitational wave induced by a particle orbiting around a Schwarzschild black hole, \href{\doibase 10.1143/PTP.90.65}{Prog. Theor. Phys. \textbf{90}, 65-84 (1993)}.

\bibitem{Cutler:1994pb}
C.~Cutler, D.~Kennefick, and E.~Poisson, Gravitational radiation reaction for bound motion around a Schwarzschild black hole, \href{\doibase 10.1103/PhysRevD.50.3816}{
Phys. Rev. D \textbf{50}, 3816-3835 (1994)}.

\bibitem{Amaro-Seoane:2012lgq}
P.~Amaro-Seoane, Relativistic dynamics and extreme mass ratio inspirals, \href{\doibase 10.1007/s41114-018-0013-8}{Living Rev. Rel. \textbf{21}, 1, 4 (2018)}, arXiv:1205.5240 [astro-ph.CO].

\bibitem{Hughes:2000ssa}
S.~A.~Hughes, Gravitational waves from extreme mass ratio inspirals: Challenges in mapping the space-time of massive, compact objects, \href{\doibase 10.1088/0264-9381/18/19/314}{Class. Quant. Grav. \textbf{18}, 4067-4074 (2001)},
arXiv:gr-qc/0008058.

\bibitem{Glampedakis:2005hs}
K.~Glampedakis, Extreme mass ratio inspirals: LISA's unique probe of black hole gravity, \href{\doibase 10.1088/0264-9381/22/15/004}{Class. Quant. Grav. \textbf{22}, S605-S659 (2005)}, arXiv:gr-qc/0509024.

\bibitem{Gair:2012nm}
J.~R.~Gair, M.~Vallisneri, S.~L.~Larson, and J.~G.~Baker, Testing General Relativity with Low-Frequency, Space-Based Gravitational-Wave Detectors, \href{\doibase 10.12942/lrr-2013-7}{Living Rev. Rel. \textbf{16}, 7 (2013)},
arXiv:1212.5575 [gr-qc].

\bibitem{Barausse:2020rsu}
E.~Barausse, E.~Berti, T.~Hertog, S.~A.~Hughes, P.~Jetzer, P.~Pani, T.~P.~Sotiriou, N.~Tamanini, H.~Witek, and K.~Yagi, \textit{et al.}, Prospects for Fundamental Physics with LISA, \href{\doibase 10.1007/s10714-020-02691-1}{Gen. Rel. Grav. \textbf{52}, 8, 81 (2020)},
arXiv:2001.09793 [gr-qc].

\bibitem{LISA:2022yao}
P.~A.~Seoane \textit{et al.} [LISA],
Astrophysics with the Laser Interferometer Space Antenna,
\href{\doibase 10.1007/s41114-022-00041-y}{Living Rev. Rel. \textbf{26}, 1, 2 (2023)},
arXiv:2203.06016 [gr-qc].

\bibitem{GWSTQLISA}
Torres-Orjuela Alejandro, Huang Shun-Jia, Liang Zheng-Cheng, Liu Shuai, Wang Hai-Tian, Ye Chang-Qing, Hu Yi-Ming, and Mei Jianwei, Detection of astrophysical gravitational wave sources by TianQin and LISA, \href{https://doi.org/10.1007/s11433-023-2308-x}{Sci. China Phys. Mech. Astron. \textbf{67}, 5, 259511 (2024)}.

\bibitem{Babak:2017tow}
S.~Babak, J.~Gair, A.~Sesana, E.~Barausse, C.~F.~Sopuerta, C.~P.~L.~Berry, E.~Berti, P.~Amaro-Seoane, A.~Petiteau, and A.~Klein, Science with the space-based interferometer LISA. V: Extreme mass-ratio inspirals,
\href{\doibase 10.1103/PhysRevD.95.103012}{Phys. Rev. D \textbf{95}, 10, 103012 (2017)}, arXiv:1703.09722 [gr-qc].

\bibitem{Zerilli:1970wzz}
F.~J.~Zerilli, Gravitational field of a particle falling in a Schwarzschild geometry analyzed in tensor harmonics,
\href{\doibase 10.1103/PhysRevD.2.2141}{Phys. Rev. D \textbf{2}, 2141-2160 (1970)}.

\bibitem{Davis:1971gg}
M.~Davis, R.~Ruffini, W.~H.~Press, and R.~H.~Price, Gravitational radiation from a particle falling radially into a Schwarzschild black hole,
\href{\doibase 10.1103/PhysRevLett.27.1466}{Phys. Rev. Lett. \textbf{27}, 1466-1469 (1971)}.

\bibitem{Davis:1972ud}
M.~Davis, R.~Ruffini, and J.~Tiomno,
Pulses of gravitational radiation of a particle falling radially into a Schwarzschild black hole,
\href{\doibase 10.1103/PhysRevD.5.2932}{Phys. Rev. D \textbf{5}, 2932-2935 (1972)}.

\bibitem{Quinn:1996am}
T.~C.~Quinn and R.~M.~Wald, An Axiomatic approach to electromagnetic and gravitational radiation reaction of particles in curved space-time,
\href{\doibase 10.1103/PhysRevD.56.3381}{Phys. Rev. D \textbf{56}, 3381-3394 (1997)}, arXiv:gr-qc/9610053.

\bibitem{Lousto:1999za}
C.~O.~Lousto, Pragmatic approach to gravitational radiation reaction in binary black holes,
\href{\doibase 10.1103/PhysRevLett.84.5251}{Phys. Rev. Lett. \textbf{84}, 5251-5254 (2000)}, arXiv:gr-qc/9912017.

\bibitem{Barack:2002mh}
L.~Barack and A.~Ori, Gravitational self-force on a particle orbiting a Kerr black hole, \href{\doibase 10.1103/PhysRevLett.90.111101}{Phys. Rev. Lett. \textbf{90}, 111101 (2003)}, arXiv:gr-qc/0212103.

\bibitem{Buonanno:1998gg}
A.~Buonanno and T.~Damour, Effective one-body approach to general relativistic two-body dynamics,
\href{\doibase 10.1103/PhysRevD.59.084006}{Phys. Rev. D \textbf{59}, 084006 (1999)}, arXiv:gr-qc/9811091.

\bibitem{Taracchini:2013rva}
A.~Taracchini, A.~Buonanno, Y.~Pan, T.~Hinderer, M.~Boyle, D.~A.~Hemberger, L.~E.~Kidder, G.~Lovelace, A.~H.~Mrou\'e, and H.~P.~Pfeiffer, \textit{et al.} Effective-one-body model for black-hole binaries with generic mass ratios and spins,
\href{\doibase 10.1103/PhysRevD.89.061502}{Phys. Rev. D \textbf{89}, 6, 061502 (2014)},
arXiv:1311.2544 [gr-qc].

\bibitem{Yunes:2009ef}
N.~Yunes, A.~Buonanno, S.~A.~Hughes, M.~Coleman Miller, and Y.~Pan,
Modeling Extreme Mass Ratio Inspirals within the Effective-One-Body Approach,
\href{\doibase 10.1103/PhysRevLett.104.091102}{Phys. Rev. Lett. \textbf{104}, 091102 (2010)},
arXiv:0909.4263 [gr-qc].

\bibitem{Zhang:2021fgy}
C.~Zhang, W.~B.~Han, X.~Y.~Zhong, and G.~Wang, Geometrized effective-one-body formalism for extreme-mass-ratio limits: Generic orbits, \href{\doibase 10.1103/PhysRevD.104.024050}{Phys. Rev. D \textbf{104}, 2, 024050 (2021)}, arXiv:2102.05391 [gr-qc].

\bibitem{Barack:2003fp}
L.~Barack and C.~Cutler, LISA capture sources: Approximate waveforms, signal-to-noise ratios, and parameter estimation accuracy,
\href{\doibase 10.1103/PhysRevD.69.082005}{Phys. Rev. D \textbf{69}, 082005 (2004)}, arXiv:gr-qc/0310125 [gr-qc].

\bibitem{Sopuerta:2011te}
C.~F.~Sopuerta and N.~Yunes, New Kludge Scheme for the Construction of Approximate Waveforms for Extreme-Mass-Ratio Inspirals, \href{\doibase 10.1103/PhysRevD.84.124060}{Phys. Rev. D \textbf{84}, 124060 (2011)}, arXiv:1109.0572 [gr-qc].

\bibitem{Chua:2017ujo}
A.~J.~K.~Chua, C.~J.~Moore, and J.~R.~Gair, Augmented kludge waveforms for detecting extreme-mass-ratio inspirals, \href{\doibase 10.1103/PhysRevD.96.044005}{Phys. Rev. D \textbf{96}, 4, 044005 (2017)},
arXiv:1705.04259 [gr-qc].

\bibitem{Liu:2020ghq}
M.~Liu and J.~d.~Zhang, Augmented analytic kludge waveform with quadrupole moment correction,
\href{https://arxiv.org/abs/2008.11396}{arXiv:2008.11396 [gr-qc]}.

\bibitem{Babak:2006uv}
S.~Babak, H.~Fang, J.~R.~Gair, K.~Glampedakis, and S.~A.~Hughes, ``Kludge'' gravitational waveforms for a test-body orbiting a Kerr black hole,
\href{https://journals.aps.org/prd/abstract/10.1103/PhysRevD.75.024005}{Phys. Rev. D \textbf{75}, 024005 (2007)}
[erratum: Phys. Rev. D \textbf{77}, 04990 (2008)], arXiv:gr-qc/0607007.

\bibitem{Ashtekar:2004eh}
A.~Ashtekar and J.~Lewandowski,
Background independent quantum gravity: A Status report, \href{\doibase 10.1088/0264-9381/21/15/R01}{
Class. Quant. Grav. \textbf{21}, R53 (2004)}, arXiv:gr-qc/0404018.

\bibitem{Perez:2017cmj}
A.~Perez, Black Holes in Loop Quantum Gravity,
\href{\doibase 10.1088/1361-6633/aa7e14}{Rept. Prog. Phys. \textbf{80}, 12, 126901 (2017)}, arXiv:1703.09149 [gr-qc].

\bibitem{Zhang:2023yps}
X.~Zhang, Loop Quantum Black Hole, \href{\doibase 10.3390/universe9070313}{Universe \textbf{9}, 7, 313 (2023)}, arXiv:2308.10184 [gr-qc].

\bibitem{Tu:2023xab}
Z.~Y.~Tu, T.~Zhu, and A.~Wang, Periodic orbits and their gravitational wave radiations in a polymer black hole in loop quantum gravity, \href{\doibase 10.1103/PhysRevD.108.024035}{
Phys. Rev. D \textbf{108}, 2, 2 (2023)}, arXiv:2304.14160 [gr-qc].

\bibitem{Liu:2024qci}
Y.~Liu and X.~Zhang, Gravitational waves for eccentric extreme mass ratio inspirals of self-dual spacetime, \href{\doibase 10.1088/1475-7516/2024/10/056}{JCAP \textbf{10}, 056 (2024)}, arXiv:2404.08454 [gr-qc].

\bibitem{Yang:2024lmj}
S.~Yang, Y.~P.~Zhang, T.~Zhu, L.~Zhao, and Y.~X.~Liu, Gravitational waveforms from periodic orbits around a quantum-corrected black hole,
\href{\doibase 10.1088/1475-7516/2025/01/091}{JCAP \textbf{01}, 091 (2025)}, arXiv:2407.00283 [gr-qc].

\bibitem{Jiang:2024cpe}
H.~Jiang, M.~Alloqulov, Q.~Wu, S.~Shaymatov, and T.~Zhu, Periodic orbits and plasma effects on gravitational weak lensing by self-dual black hole in loop quantum gravity, \href{\doibase 10.1016/j.dark.2024.101627}{Phys. Dark Univ. \textbf{46}, 101627 (2024)}.

\bibitem{Zhang:2024csc}
C.~Zhang, G.~Fu, and Y.~Gong, The constraint on modified black holes with extreme mass ratio inspirals,
\href{\doibase 10.1140/epjc/s10052-025-14100-5}{Eur. Phys. J. C \textbf{85}, 385 (2025)}, arXiv:2408.15064 [gr-qc].

\bibitem{Zi:2024jla}
T.~Zi and S.~Kumar, Eccentric extreme mass-ratio inspirals: a gateway to probe quantum gravity effects, \href{\doibase 10.1140/epjc/s10052-025-14330-7}{Eur. Phys. J. C \textbf{85}, 592 (2025)}, arXiv:2409.17765 [gr-qc].

\bibitem{Fu:2024cfk}
G.~Fu, Y.~Liu, B.~Wang, J.~P.~Wu, and C.~Zhang, Probing quantum gravity effects with eccentric extreme mass-ratio inspirals,
\href{\doibase 10.1103/PhysRevD.111.084066}{Phys. Rev. D \textbf{111}, 084066 (2025)}, arXiv:2409.08138 [gr-qc].

\bibitem{Yang:2024cnd}
S.~Yang, Y.~P.~Zhang, T.~Zhu, L.~Zhao, and Y.~X.~Liu, Constraining polymerized black holes with quasi-circular extreme mass-ratio inspirals, \href{\doibase 10.1088/1674-1137/adef1a}{Chin. Phys. C   \textbf{49}, 115107 (2025)}, arXiv:2412.04302 [gr-qc].

\bibitem{Oppenheimer:1939ue}
J.~R.~Oppenheimer and H.~Snyder,
On Continued gravitational contraction,
\href{\doibase 10.1103/PhysRev.56.455}{Phys. Rev. \textbf{56}, 455-459 (1939)}.

\bibitem{Lewandowski:2022zce}
J.~Lewandowski, Y.~Ma, J.~Yang, and C.~Zhang, Quantum Oppenheimer-Snyder and Swiss Cheese Models,
\href{\doibase 10.1103/PhysRevLett.130.101501}{Phys. Rev. Lett. \textbf{130}, 10, 101501 (2023)}, arXiv:2210.02253 [gr-qc].

\bibitem{Zhao:2024elr}
L.~Zhao, M.~Tang, and Z.~Xu,
The lensing effect of quantum-corrected black hole and parameter constraints from EHT observations,
\href{\doibase 10.1140/epjc/s10052-024-13342-z}{Eur. Phys. J. C \textbf{84}, 9, 971 (2024)},
arXiv:2403.18606 [gr-qc].

\bibitem{Ali:2024ssf}
H.~Ali, S.~U.~Islam, and S.~G.~Ghosh,
Shadows and parameter estimation of rotating quantum corrected black holes and constraints from EHT observation of M87* and Sgr A*, \href{\doibase 10.1016/j.jheap.2025.100367}{JHEAp \textbf{47}, 100367 (2025)}, arXiv:2410.09198 [gr-qc].

\bibitem{Vachher:2024ait}
A.~Vachher and S.~G.~Ghosh,
Strong gravitational lensing by rotating quantum-corrected black holes: Insights and constraints from EHT observations of M87* and Sgr A*,
\href{\doibase 10.1016/j.jheap.2024.11.012}{JHEAp \textbf{45}, 75-86 (2025)},
arXiv:2410.11332 [gr-qc].

\bibitem{Thorne:1980ru}
K.~S.~Thorne, Multipole Expansions of Gravitational Radiation,
\href{\doibase 10.1103/RevModPhys.52.299}{Rev. Mod. Phys. \textbf{52}, 299-339 (1980)}.

\bibitem{Maggiore:2007ulw}
M.~Maggiore, Gravitational Waves. Vol. 1: Theory and Experiments,
Oxford University Press, 2007.

\bibitem{Hinderer:2008dm}
T.~Hinderer and E.~E.~Flanagan, Two timescale analysis of extreme mass ratio inspirals in Kerr. I. Orbital Motion, \href{\doibase 10.1103/PhysRevD.78.064028}{
Phys. Rev. D \textbf{78}, 064028 (2008)},
arXiv:0805.3337 [gr-qc].

\bibitem{Flanagan:2012kg}
{\'E}.~E.~Flanagan, S.~A.~Hughes, and U.~Ruangsri, Resonantly enhanced and diminished strong-field gravitational-wave fluxes, \href{\doibase 10.1103/PhysRevD.89.084028}{Phys. Rev. D \textbf{89}, 8, 084028 (2014)},
arXiv:1208.3906 [gr-qc].

\bibitem{LISA:2022kgy}
K.~G.~Arun \textit{et al.} [LISA],
New horizons for fundamental physics with LISA,
\href{\doibase 10.1007/s41114-022-00036-9}{Living Rev. Rel. \textbf{25}, 1, 4 (2022)},
arXiv:2205.01597 [gr-qc].

\bibitem{gravity-book}
E. Poisson and C. M. Will, Gravity: Newtonian, Post-Newtonian, Relativistic (Cambridge University Press,
Cambridge, England, 2014).

\bibitem{Robson:2018ifk}
T.~Robson, N.~J.~Cornish, and C.~Liu, The construction and use of LISA sensitivity curves,
\href{\doibase 10.1088/1361-6382/ab1101}{Class. Quant. Grav. \textbf{36}, 10, 105011 (2019)}, arXiv:1803.01944 [astro-ph.HE].

\bibitem{Amaro-Seoane:2007osp}
P.~Amaro-Seoane, J.~R.~Gair, M.~Freitag, M.~Coleman Miller, I.~Mandel, C.~J.~Cutler, and S.~Babak, Astrophysics, detection and science applications of intermediate and extreme mass-ratio inspirals, \href{\doibase 10.1088/0264-9381/24/17/R01}{
Class. Quant. Grav. \textbf{24}, R113-R169 (2007)}, arXiv:astro-ph/0703495.

\bibitem{Amaro-Seoane:2012aqc}
P.~Amaro-Seoane, S.~Aoudia, S.~Babak, P.~Binetruy, E.~Berti, A.~Bohe, C.~Caprini, M.~Colpi, N.~J.~Cornish, and K.~Danzmann, \textit{et al.}
eLISA: Astrophysics and cosmology in the gravitational-wave millihertz regime, GW Notes \textbf{6}, 4-110 (2013), \href{https://arxiv.org/abs/1201.3621}{arXiv:1201.3621 [astro-ph.CO]}.

\bibitem{Qiao:2024gfb}
X.~Qiao, Z.~W.~Xia, Q.~Pan, H.~Guo, W.~L.~Qian, and J.~Jing, Gravitational waves from extreme mass ratio inspirals in Kerr-MOG spacetimes,
\href{\doibase 10.1088/1475-7516/2025/03/006}{JCAP \textbf{03}, 006 (2025)}, arXiv:2408.10022 [gr-qc].

\bibitem{Barack:2006pq}
L.~Barack and C.~Cutler, Using LISA EMRI sources to test off-Kerr deviations in the geometry of massive black holes,
\href{\doibase 10.1103/PhysRevD.75.042003}{Phys. Rev. D \textbf{75}, 042003 (2007)},
arXiv:gr-qc/0612029.

\bibitem{Glampedakis:2002ya}
K.~Glampedakis and D.~Kennefick, Zoom and whirl: Eccentric equatorial orbits around spinning black holes and their evolution under gravitational radiation reaction, \href{\doibase 10.1103/PhysRevD.66.044002}{Phys. Rev. D \textbf{66}, 044002 (2002)}, arXiv:gr-qc/0203086.

\bibitem{Rahman:2023sof}
M.~Rahman, S.~Kumar, and A.~Bhattacharyya, Probing astrophysical environment with eccentric extreme mass-ratio inspirals, \href{\doibase 10.1088/1475-7516/2024/01/035}{JCAP \textbf{01}, 035 (2024)}, arXiv:2306.14971 [gr-qc].

\bibitem{Zi:2025lio}
T.~Zi and S.~Kumar, Probing scalar field with generic extreme mass-ratio inspirals around Kerr black holes, \href{https://arxiv.org/abs/2508.00516}{arXiv:2508.00516 [gr-qc]}.

\bibitem{Sukova:2021thm}
P.~Sukov\'a, M.~Zaja\v{c}ek, V.~Witzany, and V.~Karas, Stellar Transits across a Magnetized Accretion Torus as a Mechanism for Plasmoid Ejection,
\href{\doibase 10.3847/1538-4357/ac05c6}{Astrophys. J. \textbf{917}, 1, 43 (2021)}, arXiv:2102.08135 [astro-ph.HE].

\bibitem{Xian:2021xkc}
J.~Xian, F.~Zhang, L.~Dou, J.~He, and X.~Shu, X-Ray Quasi-periodic Eruptions Driven by Star\textendash{}Disk Collisions: Application to GSN069 and Probing the Spin of Massive Black Holes, \href{\doibase 10.3847/2041-8213/ac31aa}{Astrophys. J. Lett. \textbf{921}, 2, L32 (2021)}, ,arXiv:2110.10855 [astro-ph.HE].

\bibitem{Pan:2021oob}
Z.~Pan, Z.~Lyu, and H.~Yang, Wet extreme mass ratio inspirals may be more common for spaceborne gravitational wave detection,
\href{\doibase 10.1103/PhysRevD.104.063007}{Phys. Rev. D \textbf{104}, 6, 063007 (2021)}, arXiv:2104.01208 [astro-ph.HE].

\bibitem{Lyu:2024gnk}
Z.~Lyu, Z.~Pan, J.~Mao, N.~Jiang, and H.~Yang, Science Opportunities of Wet Extreme Mass-Ratio Inspirals, \href{https://arxiv.org/abs/2501.03252}{
arXiv:2501.03252 [astro-ph.HE]}.

\bibitem{GRAVITY:2020gka}
R.~Abuter \textit{et al.} [GRAVITY], Detection of the Schwarzschild precession in the orbit of the star S2 near the Galactic centre massive black hole, \href{\doibase 10.1051/0004-6361/202037813}{
Astron. Astrophys. \textbf{636} (2020), L5},
arXiv:2004.07187 [astro-ph.GA].




\end{thebibliography}
\end{document}